\documentclass[onecolumn]{aastex62}

\usepackage{graphicx}
\usepackage{bm}
\usepackage{soul}
\usepackage{multirow}
\usepackage{lineno}

\graphicspath{./}

\begin{document}

\title{Evolutionary states and triplicity of  four massive semi-detached binaries with long-term decreasing orbital periods in the LMC}

\correspondingauthor{Qian Sheng-Bang}
\email{qiansb@ynu.edu.cn}

\author[0000-0002-0285-6051]{Fu-Xing Li}
\affiliation{Department of Astronomy, School of Physics and Astronomy, Yunnan University, Kunming 650091, PR China}
\affiliation{Yunnan Observatories, Chinese Academy of Sciences (CAS), P.O. Box 110, 650216 Kunming, PR China}

\author{Sheng-Bang Qian}\thanks{qiansb@ynu.edu.cn}
\affiliation{Department of Astronomy, School of Physics and Astronomy, Yunnan University, Kunming 650091, PR China}

\author{Li-ying Zhu}
\affiliation{Yunnan Observatories, Chinese Academy of Sciences (CAS), P.O. Box 110, 650216 Kunming, PR China}

\author{Wen-Ping Liao}
\affiliation{Yunnan Observatories, Chinese Academy of Sciences (CAS), P.O. Box 110, 650216 Kunming, PR China}

\author{er-gang Zhao}
\affiliation{Yunnan Observatories, Chinese Academy of Sciences (CAS), P.O. Box 110, 650216 Kunming, PR China}

\author{Min-Yu Li}
\affiliation{Yunnan Observatories, Chinese Academy of Sciences (CAS), P.O. Box 110, 650216 Kunming, PR China}

\author{Qi-Bin Sun}
\affiliation{Department of Astronomy, School of Physics and Astronomy, Yunnan University, Kunming 650091, PR China}

\author{Lin-Feng Chang}
\affiliation{Department of Astronomy, School of Physics and Astronomy, Yunnan University, Kunming 650091, PR China}

\author{Wen-Xu Lin}
\affiliation{Yunnan Observatories, Chinese Academy of Sciences (CAS), P.O. Box 110, 650216 Kunming, PR China}

\begin{abstract}
The massive semi-detached binary with a long-term decreasing orbital period may involve a rapid mass-transfer phase in Case A, and thus they are good astrophysical laboratories for investigating the evolution of massive binary stars. In this work, by using the long-term observational light curves from the OGLE project and other data in the
low-metallicity LMC,  four semi-detached massive binaries with long-term decreases in the orbital periods are detected from 165 EB-type close binaries.	It is found that the more massive component in S07798 is filling its Roche lobe where the period decrease is caused by mass transfer from the primary to the secondary. 
However, the other three (S03065, S12631, S16873) are semi-detached binaries with a lobe-filling secondary where the mass transfer between the components should cause the period to increase if the angular momentum is conservative. The long-term period decreases in these three systems may be caused by the angular momentum loss. Additionally, the orbital periods of three systems (S03065, S07798, S16873) are detected to show cyclic variation with periods shorter than 11 years, which can be plausibly explained by the presence of close-in third bodies in these massive binaries. Based on all of these
results, it is suggested that the detected four semi-detached binaries almost have multiplicity. The companion stars are crucial for the origin and evolution of these massive close binaries.
	
\end{abstract}

\keywords{binary(including multiple): close - stars: binaries: eclipsing - stars: evolution
- stars: individual (LMC)}

\section{Introduction} \label{sec:intro}
\label{sect:intro}

\label{sect:intro}

Massive stars, due to their high luminosity, dramatic evolution, and short evolutionary timescale, significantly influence the luminosity distribution, metallicity distribution, and other critical parameters of celestial systems\citep{2000ApJ...528...96P,2013ARA&A..51..457N,2023A&A...678A..60K}. Meanwhile, the massive binary is one of the most important stars, on the one hand, these binaries have a plethora of physical processes and astrophysical phenomena, such as gravitational waves\citep{2016PhRvL.116f1102A}, X-ray binaries\citep{1993ARA&A..31...93V}, gamma-ray bursts\citep{2004MNRAS.348.1215I}, etc. On the other hand, they provide an ideal test platform to determine the basic information about the components of binaries, and then provide clues to the formation and evolution that we need to understand about massive stars. Especially, since the metallicity of the Large Magellanic Cloud (LMC) is lower than that of the Milky Way \citep{2016MNRAS.455.1855C}, and massive stars have a weakness of the stellar winds at low metallicity\citep{2005A&A...443..643Y}, the study of massive binaries in the LMC may be even more significant.

In the case of massive binaries, the first point of emphasis should be their multiplicity. A study by \cite{2012Sci...337..444S} indicates that more than 70\% of massive stars exchange mass with their companions via Roche lobe overflow.  \cite{2017ApJS..230...15M} and \cite{2023ASPC..534..275O} have compiled observational and theoretical findings on stellar multiplicity, demonstrating that over 35\% of OB stars refer to triple or higher-order systems, with this proportion increasing with the component mass of the binary. \cite{2023A&A...678A..60K} discusses interactions in massive triple stars, and describes the evolution of a triple star system when taking into account the dynamics of the third body. Additionally, some results show that tertiary companions can dynamically interact with the binary, causing changes in orbital parameters through mechanisms such as Lidov-Kozai oscillations \citep{2014ApJ...793..137N,2016ARA&A..54..441N}, which suggest that tertiary companions will shed new light on the formation and evolution of massive binaries. 

Secondly, the evolution of the majority of massive binaries is also dominated by mass transfer
\citep{1999A&A...350..148W,2017ApJ...835...77M,2023MNRAS.524..471C}. \cite{2023A&A...678A..60K} shown that there are 65\%–77\% massive binaries with a companion experiencing a phase of mass transfer, initiating the so-called Case A mass transfer phase, and its donor star is often a main sequence. For instance, \cite{2023A&A...674A..56R} found a massive binary with mass transfer in this phase in the Small Magellanic Cloud. \cite{2010NewAR..54...39P} also pointed out that mass transfer occurs during the Case B process, which evolves across the Hertzsprung gap. In this process, massive binaries evolve from a detached state to a semidetached binary system, where the more massive component, fills its Roche lobe, and then the mass transfer occurs on the thermal timescale, leading to a decrease in the orbital period of the binary. However, the thermal timescale is too short, and observational samples of massive binaries are very limited. The observational results suggest that many massive binaries undergo a mass ratio inversion, leading to an increase in orbital period due to mass and angular momentum transfer\citep{2000ARA&A..38..113T,2007MNRAS.380.1599Q,2022A&A...659A..98S}. For example, TU Mus and V382 Cyg\citep{2007MNRAS.380.1599Q}, BH Cen\citep{2018RAA....18...59Z}, V Pup\citep{2008ApJ...687..466Q,2021MNRAS.502.6032B}, etc, represent this class of massive binaries with a long-term increasing orbital period. We also try to find those massive binaries with long-term decreasing orbital periods, but these observational samples are rare, with V606 Cen\citep{2022ApJ...924...30L} and GU Mon\citep{2019AJ....157..111Y} serving as representatives. Therefore, this study aims to search for these rare massive binaries in the LMC, focusing on semidetached binaries, which provide favourable conditions for studying mass transfer in massive stars. This will provide observational evidence to test theories on the formation and evolution of massive binaries.

Based on the Optical Gravitational Lensing Experiment (OGLE) project, it is an ideal laboratory for providing numerous light curves and long-term continuous observations, \cite{2016AcA....66..421P} discovered 40\,204 EBs in the LMC from the fourth phase of the OGLE project, and a large number of semidetached binary candidates can be obtained from the non-contact binaries in this catalog. Moreover, substantial results have been obtained on the period variations in binary systems\citep{2017MNRAS.472.2241Z,2019MNRAS.485.2562H,2022MNRAS.509..246H}, which suggest that it is a valuable opportunity to search for the targets of this study. Through detailed studies of semidetached massive binaries with long-term decreasing orbital periods, it has been possible to explore the formation channels and evolutionary paths of massive binaries from the discussion of mass transfer, mass and angular momentum loss, and the influence of third bodies.
The outline of the paper is as follows: Section 2 explains the source of the data acquisition and the massive binary systems obtained, Section 3 provides a comprehensive analysis of the changes in orbital periods, Section 4 accomplishes the determination of the physical parameters, and Section 5 offers the discussion and conclusion.

\begin{table}
\caption{The basic information of the four binaries.}
\begin{center}
\begin{tabular}{llllccc}\hline\hline
Parameters	&	S03065	&	S07798	&	S12631	&	S16873	\\	\hline
Period(days)	&	1.0801861	&	0.8004206	&	1.0702108	&	1.0793907		\\
Epoch(HJD-2\,450\,000)	&	7\,000.2805	&	7\,000.0085	&	7\,000.4173	&	7\,000.1451		\\
$I$(mag)	&	15.961	&	17.009	&	16.071	&	16.07		\\
$V$(mag)	&	15.788	&	16.937	&	15.915	&	15.948		\\
$V-I$(mag)	&	-0.173	&	-0.072	&	-0.156	&	-0.122		\\
$E(V-I)$(mag)	&	0.068	&	0.109	&	0.078	&	0.077		\\
$(V-I)_0$(mag)	&	-0.241	&	-0.181	&	-0.234	&	-0.199		\\
$T_{M_v}$(K)	&	21\,500	&	16\,600	&	20\,900	&	17\,600		\\
Source	&	OGLE III+IV, TESS	&	OGLE III+IV+II	&	OGLE III+IV, EROS-2, TESS	&	OGLE III+IV+II, TESS		\\
\hline
\end{tabular}
\end{center}
\label{table:1}
\end{table}

\begin{figure}
\begin{center}
\includegraphics[width=60mm,height=60mm]{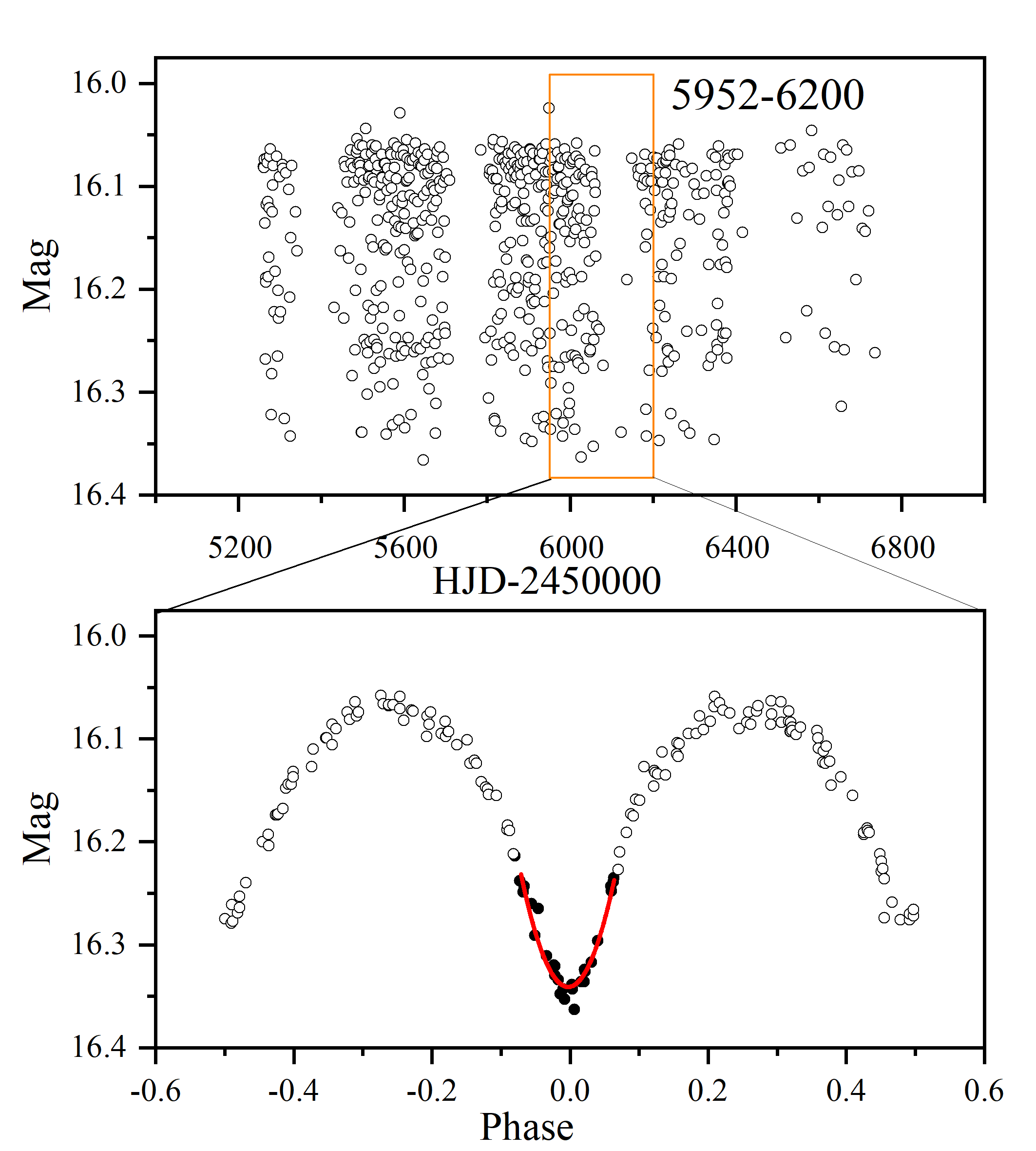}
\caption{A eclipse time of S12631. The lower panel shows the phase curves obtained from converting 120 data points. A parabolic fit was performed using the data within a phase width of 0.14 to determine the time of the minimum. }
\end{center}
\label{figure:1}
\end{figure}

\section[]{DATA AND BINARY SYSTEMS}
\subsection{Data source} \label{sec:data}

The OGLE project is a long-term project with the main goal of searching for the dark matter with microlensing phenomena. However, it has accumulated much valuable data with long-term observations over the past 30 years in specific regions, including light curves of eclipsing binaries in the Magellanic Clouds and the Milky Way. These light curves have stimulated our interest in studying the evolution of massive binaries in the LMC. Therefore, we downloaded all the light curves from the OGLE\footnote{http://ogle.astrouw.edu.pl/} in the LMC, there are mainly two parts (OGLE III and OGLE IV), and these curves were obtained with the $I$ and $V$ band filters, and took about 90\% data in the $I$ band. So, only the $I$ light curves were used to drive the orbital period changes, and those of both bands were used to determine the fundamental parameters for the binaries. As well as using OGLE III and OGLE IV, some datasets and survey data have also been used to extend the time span of the observed targets. These include the OGLE II and EROS-2 surveys\footnote{https://vizier.cds.unistra.fr/viz-bin/VizieR}, and data from the Transiting Exoplanet Survey Satellite 
(TESS; \citealt{2015JATIS...1a4003R}), that have been obtained with high-quality aperture photometry. EROS-2 and OGLE II conducted observations almost simultaneously, which is used when OGLE II is not available. The detailed information is given in Table 1. The OGLE project was equipped with the chip mosaic CCD camera on the 1.3-meter diameter telescope at Las Campanas Observatory, Chile. The EROS-2 survey was carried out with the 1-meter telescope at ESO, La Silla.

TESS is a great all-sky survey that is designed to detect small transiting planets orbiting the brightest and nearest stars in the sky\citep{2018AJ....156..102S}. The targets in the LMC have been observed many times in the TESS project, sector 1-13, sector 27-39, and sector 61-69. Since the official did not include light curves for these four targets, the light curves were extracted by applying a new aperture photometry program to the TESScut \citep{2019ascl.soft05007B} fits\footnote{https://mast.stsci.edu/tesscut/}, which are available at MAST\dataset[doi:10.17909/r1b8-aj60]{https://archive.stsci.edu/doi/resolve/resolve.html?doi=10.17909/r1b8-aj60}. When extracting the light curves, the percentile threshold of background and aperture were set to be 10\% and 70\%, respectively, only one pixel was selected as the aperture and the detrending was performed using the Something method (Lowess). Despite our efforts to obtain light curve data from all sectors, some sectors were excluded due to poor data quality. As a result, due to the dense star field and the fact that these candidate binaries are not very bright, the TESS light curves have a larger scatter than those of OGLE. We only used sectors with better data quality, and one of the targets was not derived from the TESS light curves. However, the TESS data will be useful for further analysis and study.

\subsection{Targets} \label{sec:data}

The aim of this study is to study the evolution of the massive semidetached binaries with long-term orbital period decreases in the LMC. Therefore, several methods have been used to find these binaries, and the detailed information is as follows: (1) Based on the work from \cite{2016AcA....66..421P}, which listed the classification (contact and non-contact binaries), and all non-contact binaries were selected. (2) This work focused on the massive binary systems (i.e., spectral type earlier than B5V after considering reddening), we used the criteria with a color index of $V-I$ below $0.0$ and $(V-I)_0$ below $-0.165$. The basic information of the binaries were derived from \cite{2016AcA....66..421P}, and $E(V-I)$ values were provided by \cite{2021ApJS..252...23S}.These binaries were removed when the eclipse depth was below 0.1 magnitudes for high-quality results, and there were 1178 targets left. (3) We got the binaries with EB-type or EB-like light curves by eyes, and 165 candidates were determined. (4) We created the O-C (observational minima - calculated minima) curves based on the minima of the binaries for these 165 candidates using OGLE III and IV light curves. Then the desired result has been obtained with a downward parabolic trend in the O-C curves, and the trend may imply a long-term decrease in the orbital period and result in about 10 systems being retained. (5) The time series is very important to determine a change of orbital period. So, the data from OGLE II, EROS-2, and TESS were used to identify and test the changes for these massive binaries. Finally, by deriving the light curve solutions, 4 semidetached massive binaries with a long-term decrease in orbital period are identified. The details are given in Table \ref{table:1}. For convenience,  all of the systems are referred to as SXXXXX from the initial name of OGLE LMC-ECL-XXXXX in this work.

\begin{figure}
\begin{center}
\includegraphics[width=60mm,height=60mm]{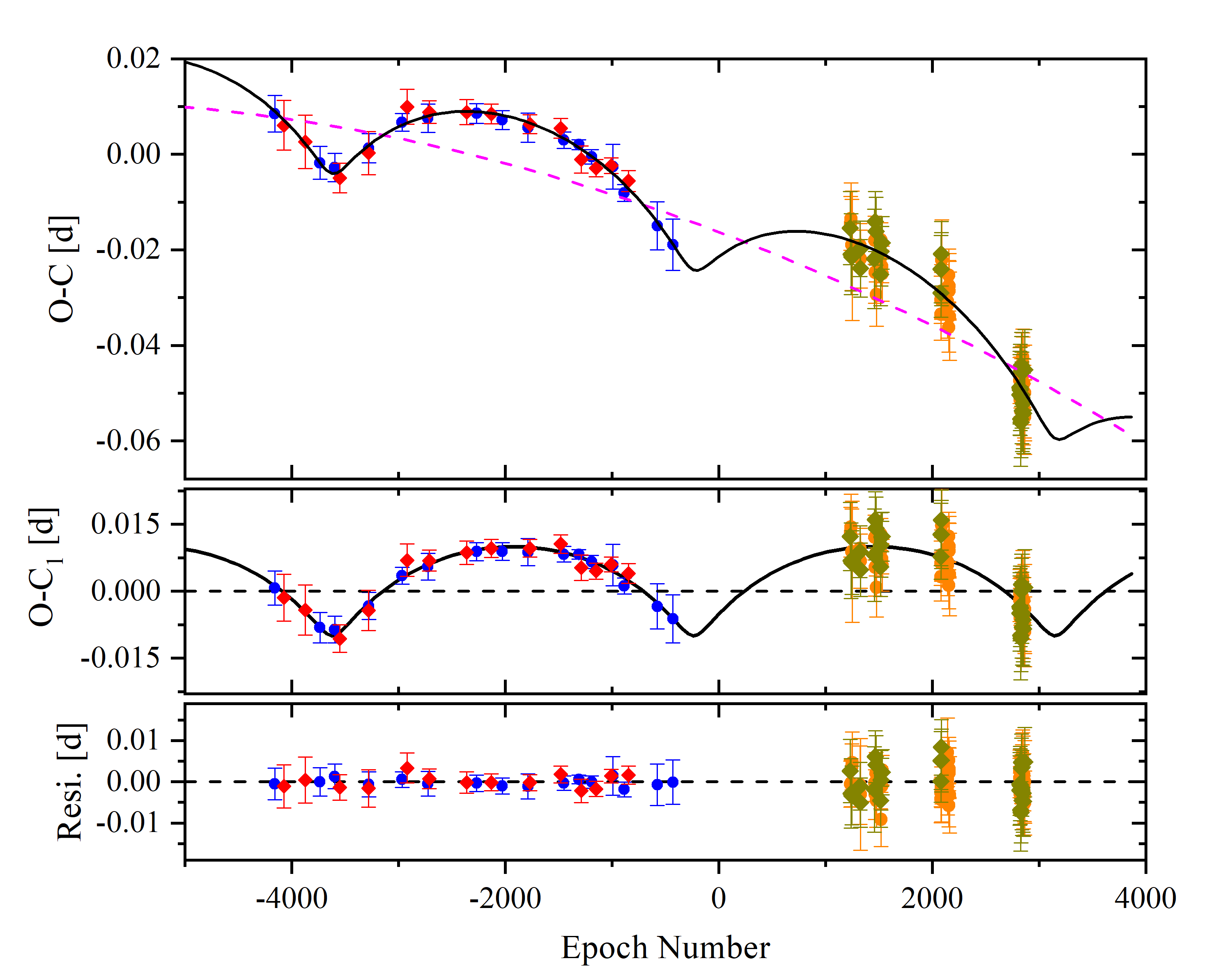}
\includegraphics[width=60mm,height=60mm]{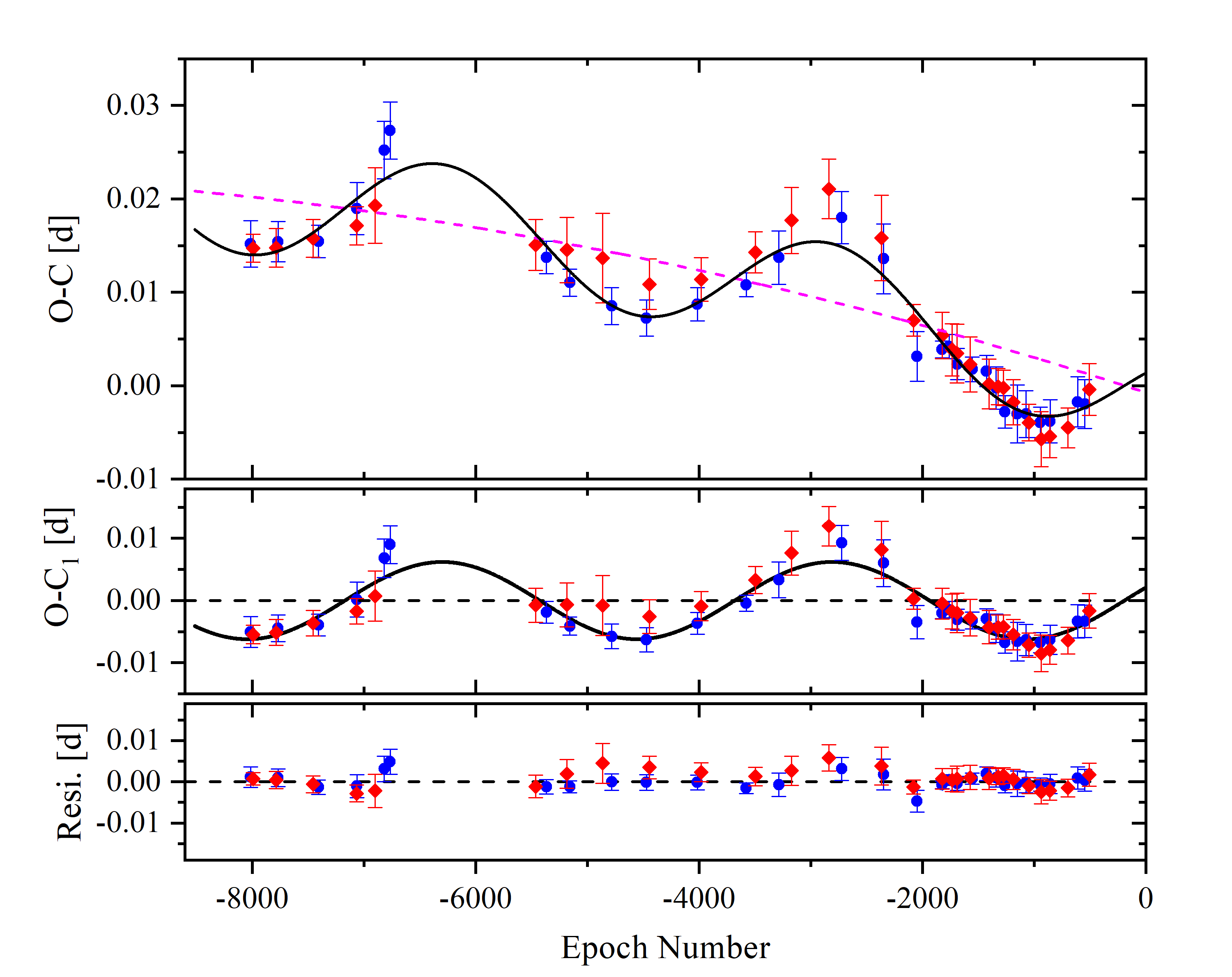}\\
S03065 \hspace{16em} S03065\\
\includegraphics[width=60mm,height=60mm]{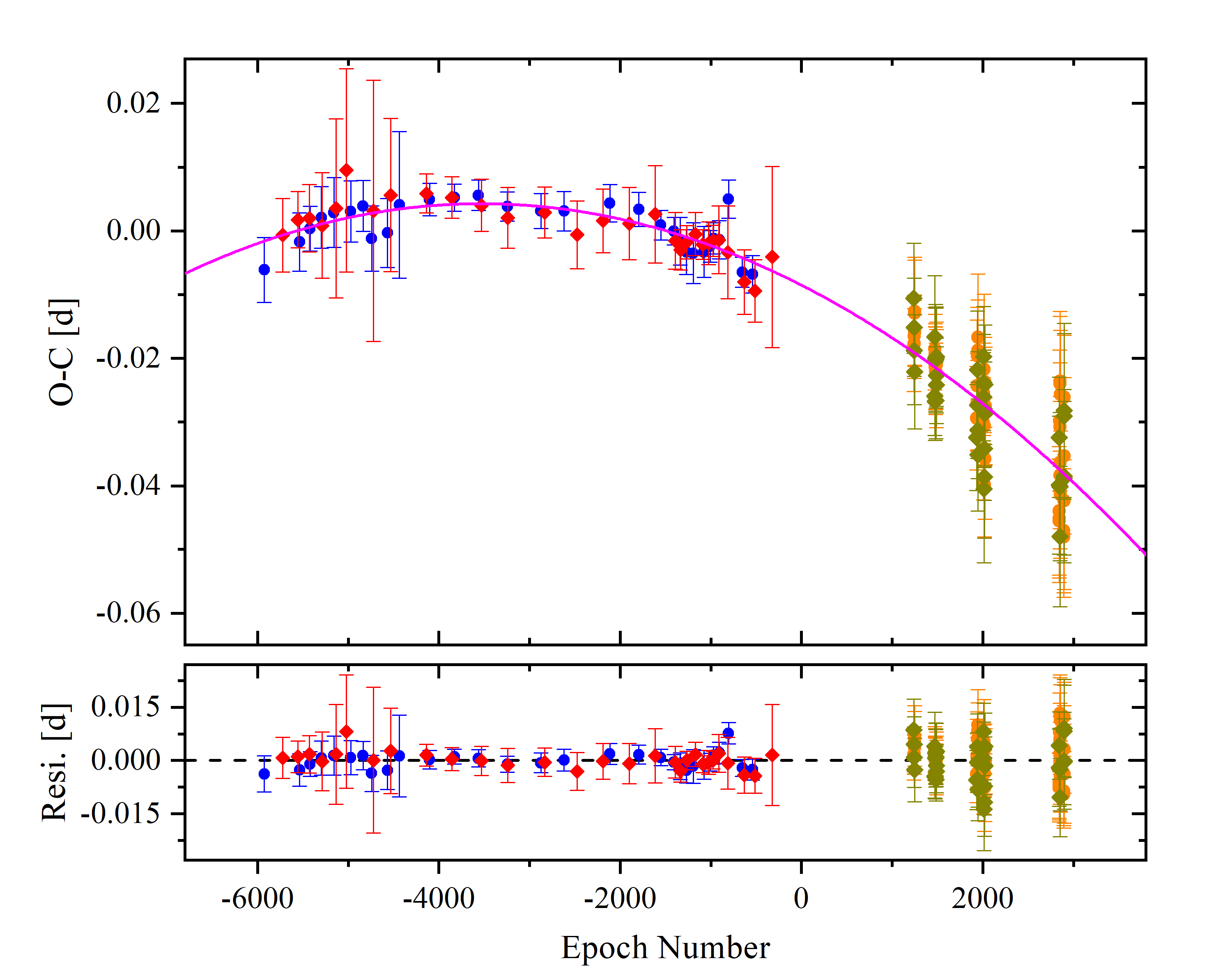}
\includegraphics[width=60mm,height=60mm]{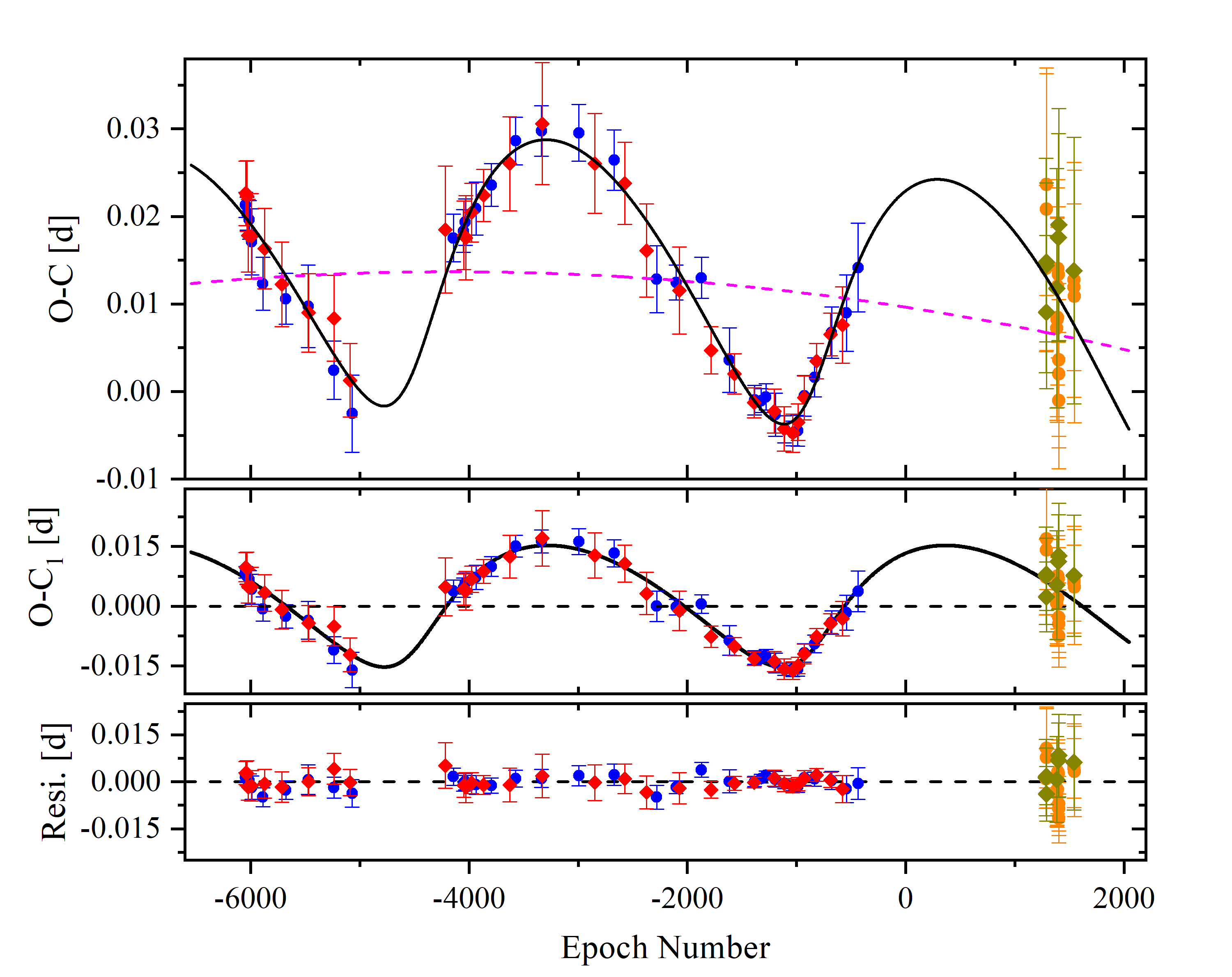}\\
S12631 \hspace{16em} S16873\\
\caption{The (O$-$C) diagrams for four binaries. The upper panels represent the (O$-$C) curves using all minima (the blue and red dots derived from OGLE or EROS-2, the orange and dark yellow symbols obtained from TESS), while the lines refer to their trends. The middle panels (except for S12631) and lower panels represent the cyclic oscillations and the residuals respectively.}
\end{center}
\label{figure:oc}
\end{figure}

\section{PERIOD VARIATIONS AND LIGHT TRAVEL-TIME EFFECT} \label{sec:ltte}

 \subsection{Eclipse times} \label{sec:data}

In binary systems, the time of a minimum (eclipsed by its component) provides valuable information about the system. Analysing the variations of these minima (O-C) based on the theory of light travel$-$time effect (LTTE), reveals physical phenomena such as changes in magnetic activity or the influence of a third body. For instance, a periodic change represents the gravitational effect of a third body orbiting the binary system\citep{2022ApJ...924...30L,2024ApJ...967...81L}. Additionally, mass transfer between components or the loss of mass and angular momentum can lead to parabolic changes in the O-C curve\citep{2007MNRAS.380.1599Q,2021AJ....162...13L}.

In this study of massive binaries, periodic oscillations in O-C are typically attributed to gravitational perturbations from a third body. For the O-C analysis, more sufficient minima have been used, making the results more accurate. Traditionally, an eclipse timing is obtained from continuous observations during a primary or secondary eclipse. However, with the release of large survey data, a new approach to obtaining minima has been adopted. This method converts discontinuous time series into phases and derives a minimum by shifting the phase over a period, and all the minima in this paper have been derived using survey data by this method. \cite{2022MNRAS.514.1206L} has confirmed the reliability of this approach, which is similar to the method employed by \cite{2014A&A...572A..71Z} and referred to as semi-automatic fitting procedure (APF). This has led to the discovery of many hierarchical triple system candidates in the Galactic Bulge \citep{2019MNRAS.485.2562H,2022MNRAS.509..246H} and the Small Magellanic Cloud \citep{2017MNRAS.472.2241Z}.

We employed 120 data points to determine the primary and secondary minima (as illustrated in Figure 1) for OGLE and EROS-2 in this work. Then 50 data points were shifted to determine the subsequent minimum, which can obtain more and denser minima. The continuous light curves obtained from TESS data have a large dispersion, and the minima used in the text are also fitted using the same method with data of multiple periods. The TESS times of minima were originally determined in BJD and then converted to HJD. All the minima of four massive binaries are listed in the table in the appendix. 

 \subsection{Period variation analysis} \label{sec:data}

Based on the minima obtained for four massive binaries, we use the LTTE to perform the O-C calculation. For this step, correct linear ephemerides are required, which are taken from \cite{2016AcA....66..421P} as detailed in Table \ref{table:1}. For these four massive binaries, the time spans range from 16.5 to 25.9 years, and the O-C curves are plotted in the upper panel of Figure \ref{figure:oc}. In these figures, the O-C values for primary and secondary minima are displayed in different colours, the primary minima for OGLE and EROS-2 are represented in blue, while that for TESS are shown in orange. The primary minima are calculated using the linear ephemerides from Table \ref{table:1}.

The O-C curves indicate that all four targets require fitting with downward parabolas, suggesting a long-term orbital decrease, which precisely aims to identify in such special massive binary systems, with S12631 exhibiting this behaviour prominently. Additionally, three out of the four show periodic oscillations, with only S12631 lacking this characteristic, and the strictly periodic variations exceed two cycles. All four binaries are early-type spectroscopic binaries, the most reasonable explanation involves the presence of a third body orbiting the central binary system, which was created by the LTTE. Although periodic variations may also be caused by magnetic activity and mass motion in late-type binaries.

During the fitting process, S03065 and S168731 were fitted with eccentric orbits, while S07798 was fitted with a circular orbit, it is noted that the least-square fit with weights of the errors was used as described in Equation 1.
\begin{equation}
    O-C = \Delta{T_0} +\Delta{P_0} \times E + \beta \times E^2 + \tau,
\end{equation}
where E represents the epoch number, ${\beta}$ indicates the rate of the linear period change, $\Delta{T_0}$ and $\Delta{P_0}$ denote the updated epoch and period respectively, $\tau$ is the cyclic modulation term induced by the LTTE, which represents the sinusoidal or eccentric function for these three binaries. The eccentric function can be referenced from \cite{2022MNRAS.514.1206L}, which provides detailed mathematical equations. We obtain the periodic modulation of O-C after fitting, which allows us to derive parameters of the third body, such as semi-amplitude, period of the third body, and other related parameters. The corresponding parameters and the revised values are listed in Table \ref{table:3d}. Where f(m) denotes the mass function of systems hosting a third body, A and e refer to the semi-amplitude and eccentricity respectively. $M_3\sin({i_{3}=90^{\circ}})$ is the lowest mass of the third body, determined from the mass of the binaries. $D_{max}$ is the maximum distance between the third body and the binary, determined using Kepler's third law, the observed orbital period ($P_3$) and the mass of the binary. It should be noted that due to the temporal smoothing associated with the determination of the minima, we perform a smaller semi-amplitude than the actual one, potentially leading to a slight underestimation of the third body parameters derived from the value of the semi-amplitude.

\begin{table}
\caption{Orbital parameters of the third body or revised period for four massive binaries.}
\begin{center}
\begin{tabular}{llllccc}\hline\hline
Parameters	&	S03065	&	S07798	&	S12631	&	S16873	\\	\hline
$\Delta{T_{0}}$(d)$\times{10^{-3}}$	&	-16.3($\pm$1.1)	&	-0.7($\pm$1.0)	&	-8.5($\pm$0.3)	&	9.6($\pm$0.8)	\\	
$\Delta{P_{0}}$(d)$\times{10^{-6}}$	&	-8.49($\pm$0.20)	&	-3.93($\pm$0.57)	&	 -7.24($\pm$0.17)	&	-1.94($\pm$0.48)	\\	
$\beta\times{10^{-10}}({d \,cycle^{-1}})$	&	-6.53($\pm$0.98)	&	-1.63($\pm$0.63)	&	-10.25($\pm$0.44)	&	-2.34($\pm$0.74)	\\	
The semi-amplitude, A(d)$\times{10^{-3}}$	&	10.0($\pm$0.9)	&	6.2($\pm$0.6)	&	\nodata	&	16.0($\pm$0.7)	\\	
Orbital period, $P_{3}$ (yr)	&	10.01($\pm$0.13)	&	7.64($\pm$0.13)	&	\nodata	&	10.76($\pm$0.10)	\\	
Orbital phase, $\varphi$(degrees)	&	\nodata	&	20.0($\pm$5.7)	&	\nodata	&	\nodata	\\	
Longitude of the periastron passage, $\omega$(degrees)	&	269.7($\pm$8.8)	&	\nodata	&	\nodata	&	315.3($\pm$8.2)	\\	
Periastron passage, $T_{3}$(HJD)	&	2\,459\,739.43($\pm$94.22)	&	\nodata	&	\nodata	&	2\,460\,445.11($\pm$73.05)	\\	
Eccentricity, e	&	0.69($\pm$0.11)	&	0	&	\nodata	&	0.42($\pm$0.04)	\\	
Mass function, f(m)($M_{\odot}$)	&	0.052($\pm$0.010)	&	0.021($\pm$0.006)	&	\nodata	&	0.213($\pm$0.027)	\\	
Projected semimajor axis, $a_{12}\sin{i_{3}}$ (au)	&	1.73($\pm$0.16)	&	1.07($\pm$0.10)	&	\nodata	&	2.91($\pm$0.12)	\\	
Projected masses, $M_3\sin{i_{3}}$($M_{\odot}$) 	&	1.90($\pm$0.09)	&	1.19($\pm$0.03)	&	\nodata	&	2.94($\pm$0.08)	\\	
$D_{max}$(au)	&	11.23($\pm$1.77)	&	8.04($\pm$1.18)	&	\nodata	&	10.81($\pm$1.20)	\\

\hline
\end{tabular}
\end{center}
\label{table:3d}
\end{table}

\begin{table}
\caption{Photometric solutions for four semidetached massive binaries.}
\begin{center}
\begin{tabular}{llllccccc}\hline\hline
Parameters	&			S03065	&		&	S07798	&		&	S12631	&		&	S16873	\\	\hline
Mode(Semi-detached)	
& \multicolumn{1}{@{}c@{}}{5}& \multicolumn{3}{@{}c@{}}{4} & \multicolumn{1}{@{}c@{}}{5} & \multicolumn{3}{@{}c@{}}{5}\\

Case	&	no $L_{3}$	&	$L_{3}$	&	no $L_{3}$	&	$L_{3}$	&	no $L_{3}$	&	$L_{3}$	&	no $L_{3}$	&	$L_{3}$	\\
$q$($M_{2}$/$M_{1}$)	&	$0.32^{+0.05}_{-0.04}$	&	$0.32^{+0.02}_{-0.02}$	&	$0.49^{+0.01}_{-0.02}$	&	$0.49^{+0.01}_{-0.01}$	&	$0.29^{+0.11}_{-0.08}$	&	$0.30^{+0.01}_{-0.01}$	&	$0.40^{+0.13}_{-0.09}$	&	$0.41^{+0.01}_{-0.01}$	\\	
$i$($^{\circ}$)	&	$74.70^{+1.12}_{-1.13}$	&	$78.00^{+1.10}_{-1.10}$	&	$82.21^{+0.54}_{-0.51}$	&	$84.41^{+0.42}_{-0.42}$	&	$66.70^{+1.77}_{-1.40}$	&	$67.26^{+0.66}_{-0.66}$	&	$64.86^{+1.04}_{-0.81}$	&	$67.80^{+0.33}_{-0.33}$	\\	
$T_2$(K)	&	$14212^{+168}_{-157}$	&	$14018^{+163}_{-163}$	&	$11172^{+79}_{-82}$	&	$10804^{+75}_{-75}$	&	$15504^{+299}_{-271}$	&	$15508^{+115}_{-115}$	&	$13178^{+234}_{-231}$	&	$12962^{+74}_{-74}$	\\	
$r_{1,2}$	&	$0.43^{+0.01}_{-0.01}$	&	$0.435^{+0.003}_{-0.003}$	&	$0.27^{+0.01}_{-0.01}$	&	$0.277^{+0.004}_{-0.004}$	&	$0.46^{+0.03}_{-0.03}$	&	$0.461^{+0.002}_{-0.002}$	&	$0.42^{+0.02}_{-0.03}$	&	$0.433^{+0.008}_{-0.008}$	\\	
$\Omega_{1,2}$	&	$2.745^{+0.016}_{-0.016}$	&	$2.707^{+0.043}_{-0.043}$	&	$3.101^{+0.008}_{-0.008}$	&	$3.074^{+0.033}_{-0.033}$	&	$2.542^{+0.008}_{-0.008}$	&	$2.573^{+0.019}_{-0.019}$	&	$2.863^{+0.008}_{-0.008}$	&	$2.816^{+0.010}_{-0.010}$	\\	
$L_{1}/(L_{1}+L_{2})$(V)	&	\nodata	&	\nodata	&	$0.859^{+0.003}_{-0.003}$	&	$0.851^{+0.003}_{-0.003}$	&	$0.827^{+0.005}_{-0.005}$	&	$0.819^{+0.005}_{-0.005}$	&	$0.762^{+0.006}_{-0.006}$	&	$0.776^{+0.003}_{-0.003}$	\\	
$L_{1}/(L_{1}+L_{2})$(I)	&	$0.808^{+0.009}_{-0.009}$	&	$0.817^{+0.007}_{-0.007}$	&	$0.847^{+0.003}_{-0.003}$	&	$0.839^{+0.003}_{-0.003}$	&	$0.817^{+0.005}_{-0.005}$	&	$0.809^{+0.005}_{-0.005}$	&	$0.750^{+0.006}_{-0.006}$	&	$0.764^{+0.003}_{-0.003}$	\\	
$L_{3}/(L_{1}+L_{2}+L_{3})$($V\%$)	&	\nodata	&	\nodata	&	\nodata	&	$2.0^{+1.9}_{-1.9}$	&	\nodata	&	$2.8^{+3.4}_{-2.8}$	&	\nodata	&	$16.4^{+1.4}_{-1.4}$	\\	
$L_{3}/(L_{1}+L_{2}+L_{3})$($I\%$)	&	\nodata	&	$11.3^{+3.2}_{-3.2}$	&	\nodata	&	$3.6^{+1.8}_{-1.8}$	&	\nodata	&	$5.6^{+3.2}_{-3.2}$	&	\nodata	&	$16.8^{+1.3}_{-1.3}$	\\	
f($\%$)	&	$70.8^{+0.9}_{-0.9}$	&	$74.2^{+0.3}_{-0.3}$	&	$61.6^{+0.5}_{-0.5}$	&	$65.1^{+2.9}_{-2.9}$	&	$86.4^{+0.7}_{-0.7}$	&	$85.7^{+1.7}_{-1.7}$	&	$77.0^{+0.5}_{-0.5}$	&	$84.4^{+0.8}_{-0.8}$	\\	\hline

\end{tabular}
\end{center}
\label{table:timescale}
\end{table}

\begin{table}
\caption{The timescale for four semidetached massive binaries.}
\begin{center}
\begin{tabular}{llllccccc}\hline\hline
Parameters	&	S03065	&	S07798	&	S12631	&	S16873	\\	\hline
$M_1$($M_{\odot}$)	&	$7.9^{+1.6}_{-1.6}$	&	$5.2^{+1.0}_{-1.0}$	&	$7.5^{+1.5}_{-1.5}$	&	$5.7^{+1.1}_{-1.1}$	\\	
$L_1$($L_{\odot}$)	&	$3291^{+995}_{-995}$	&	$696^{+215}_{-215}$	&	$3205^{+1320}_{-1320}$	&	$1206^{+451}_{-451}$	\\	
$\dot{P}({d \,yr^{-1}})\times10^{-7}$ 	&	$-4.41^{+0.66}_{-0.66}$	&	$-1.49^{+0.62}_{-0.62}$	&	$-7.00^{+0.30}_{-0.30}$	&	$-1.58^{+0.50}_{-0.50}$	\\	
$\dot{M_1}$($M_{\odot}$)	&	$-5.06^{+1.32}_{-1.32}$	&	$-3.10^{+1.42}_{-1.42}$	&	$-6.68^{+1.74}_{-1.74}$	&	$-1.85^{+0.70}_{-0.70}$	\\	
$\tau$ $(yr)\times10^{7}$ 	&	$0.50^{+0.13}_{-0.13}$	&	$1.48^{+0.77}_{-0.77}$	&	$0.33^{+0.08}_{-0.08}$	&	$1.23^{+0.46}_{-0.46}$	\\	
$\tau_\textrm{th}  (yr)\times10^{5}$ 	&	$2.56^{+0.78}_{-0.78}$	&	$3.79^{+2.01}_{-2.01}$	&	$1.59^{+0.80}_{-0.80}$	&	$3.07^{+1.31}_{-1.31}$
	\\	\hline

\end{tabular}
\end{center}
\label{table:time}
\end{table}

\section{LIGHT-CURVE MODELING} \label{subsec:wd}

Binaries are important because their light curves can be used to determine some of the basic physical parameters of the two components in a binary system. Initially, tools such as the Wilson and Devinney (W-D) program\citep{1971ApJ...166..605W,1990ApJ...356..613W,2003ASPC..298..323V} were used for this purpose, but now the PHysics Of Eclipsing BinariEs (PHOEBE)\footnote{https://phoebe-project.org/} program has attracted more attention. In this work, we modeled the light curves for these binaries using the V, I band data from OGLE IV. With the exception of S03065, only I-band data were observed for this binary. For modeling the light curves of these binary stars, we first used the 2013 version of the W-D program to obtain fundamental parameters and to preliminarily determine the geometric structures of these binaries. Subsequently, to further confirm the results and enhance their reliability, we employed the MCMC method using PHOEBE (2.4.11 script) for modeling and validation\citep{2005ApJ...628..426P,2023MNRAS.525.4596D,2023RAA....23i5011P}. To conserve computational resources, we have used only a portion of the OGLE IV data (within HJD 24\,560\,000) for the PHOEBE calculation. Although this is partial data, it still covers many periods, and the results of the W-D solution show that the partial data results for these four target stars are almost identical to those obtained using the full OGLE IV data. Similarly, the results of PHOEBE are also similar to those of W-D.

\begin{figure}
\begin{center}
\includegraphics[width=60mm,height=60mm]{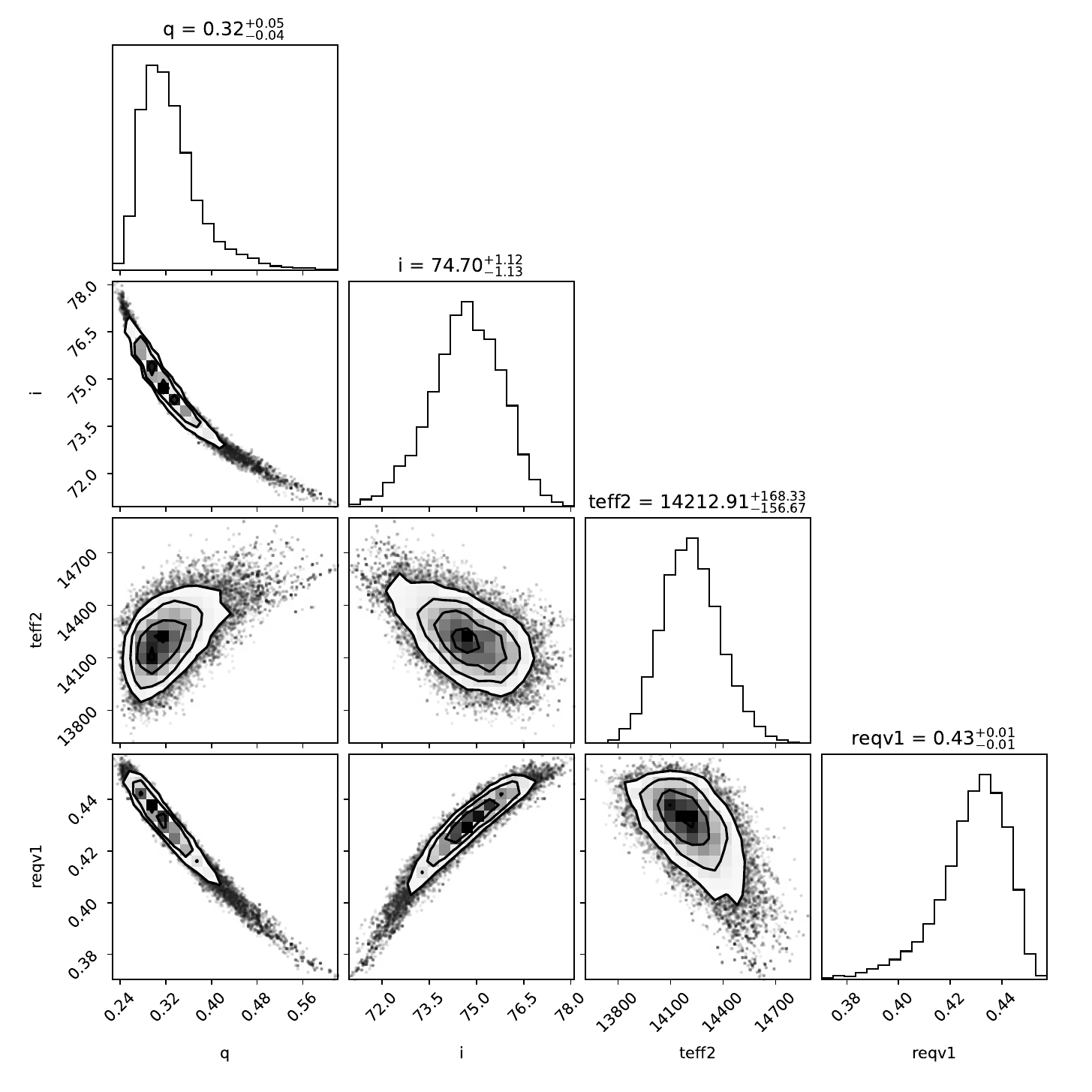}
\includegraphics[width=60mm,height=60mm]{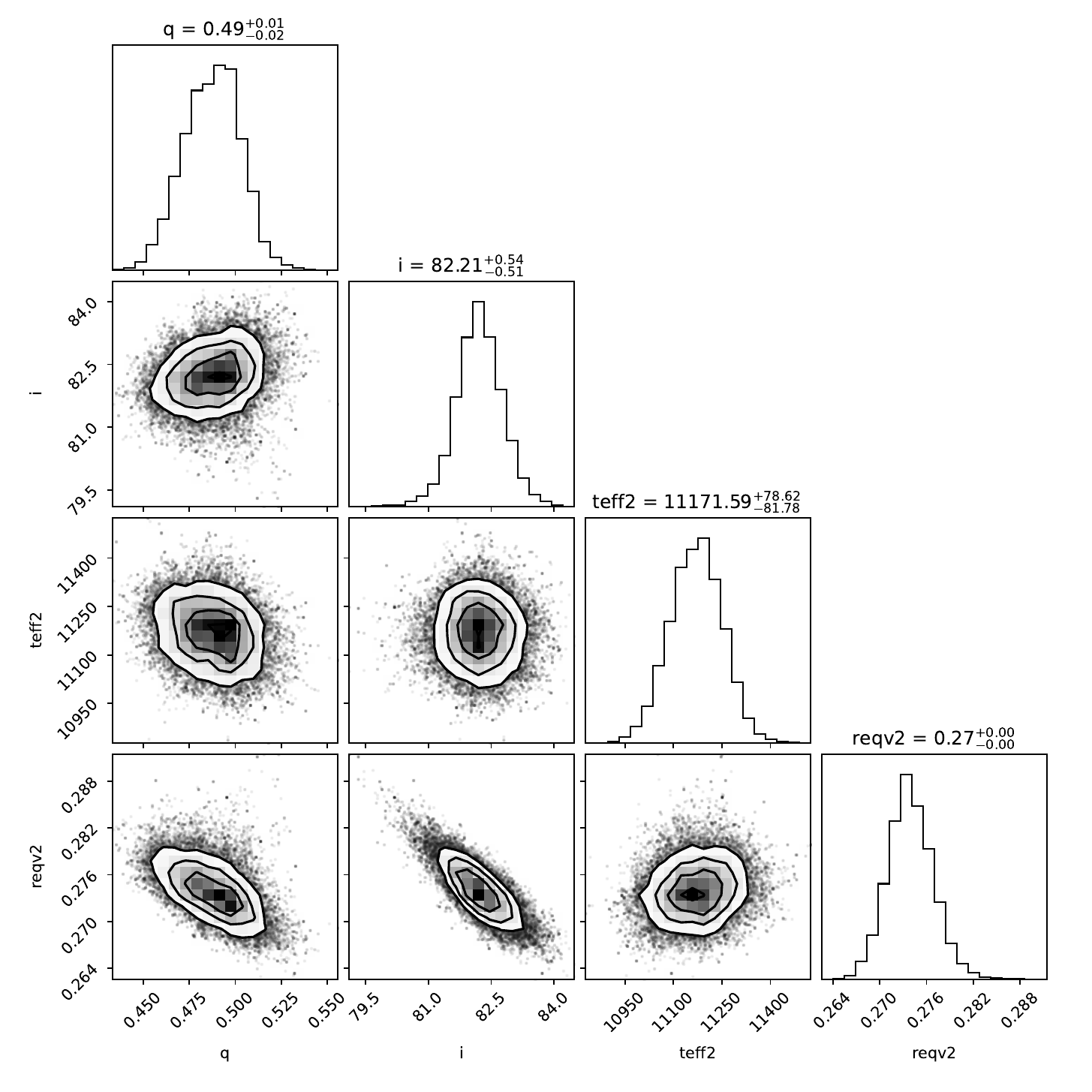}\\
S03065 \hspace{16em} S03065\\
\includegraphics[width=60mm,height=60mm]{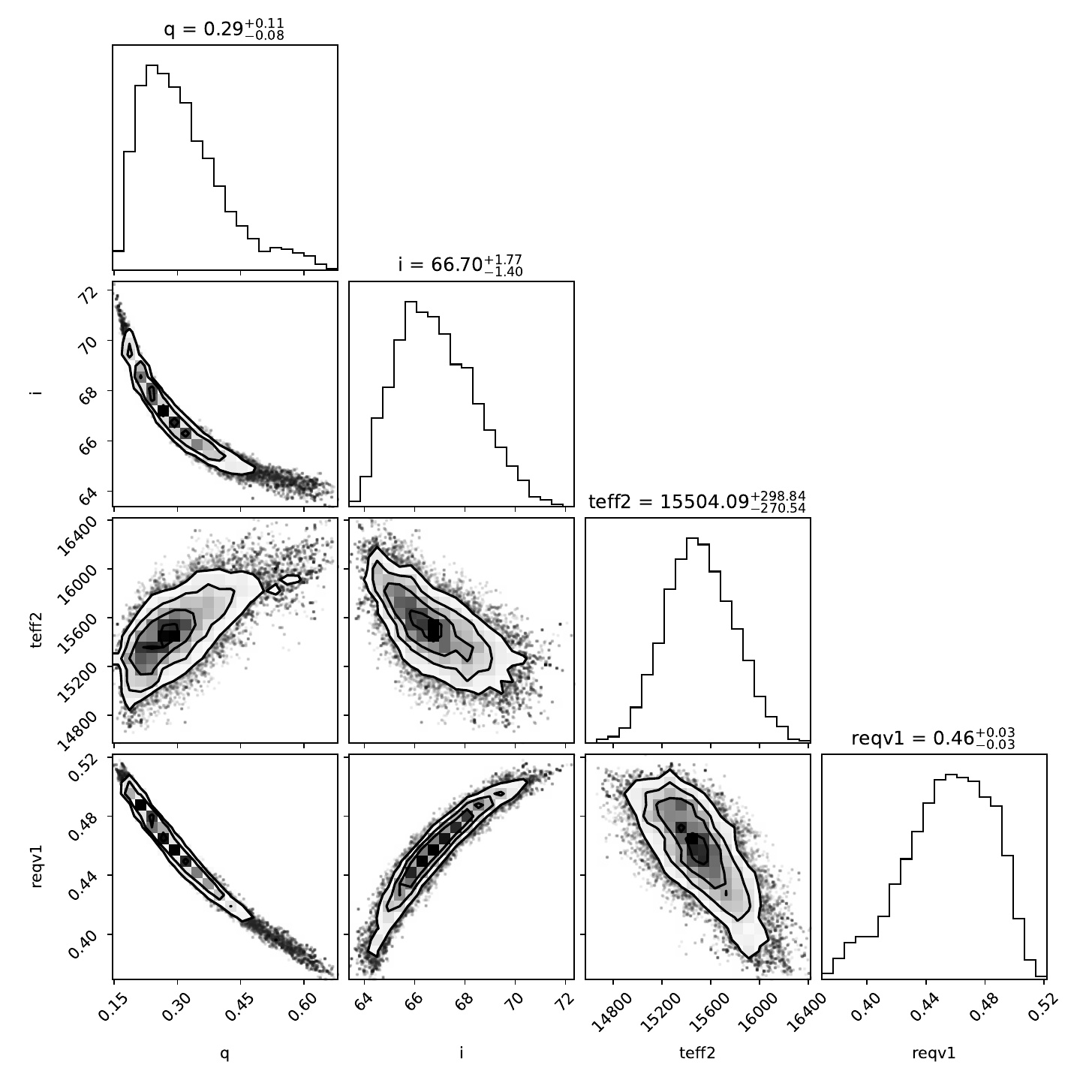}
\includegraphics[width=60mm,height=60mm]{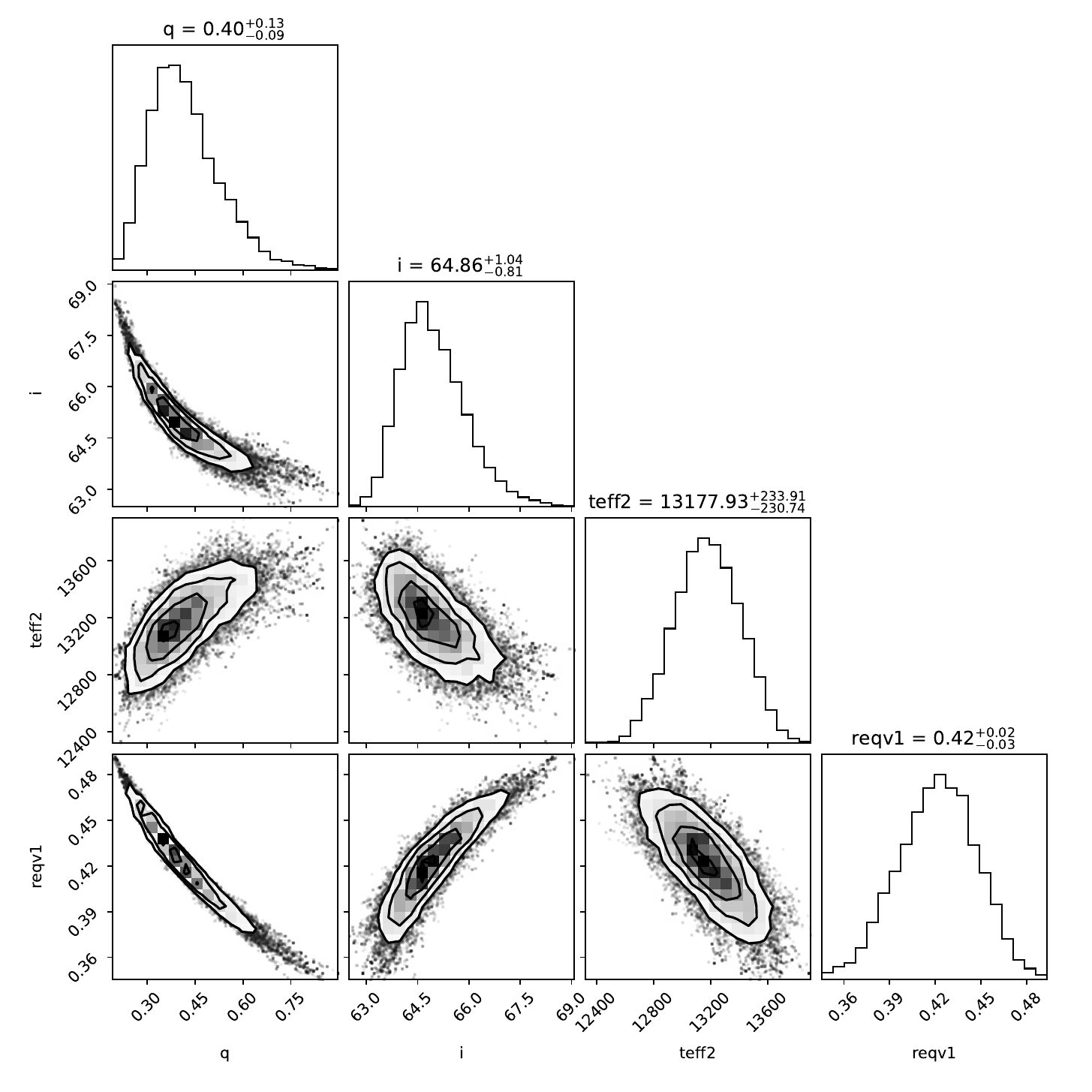}\\
S12631 \hspace{16em} S16873\\
\caption{The distribution of the parameters for four massive binaries from the PHOEBE program. q, i, teff2, reqv refer to the mass ratio, orbital inclination, temperature of the secondary star, and the radius of the component that has not filled its Roche lobe, respectively.}
\end{center}
\label{figure:pb}
\end{figure}

Before light curve modeling, it is necessary to determine some initial parameters that need to be input, such as temperature,  gravity-darkening coefficients, and bolometric albedo. These binaries lack spectroscopy, making it difficult to determine accurate temperatures. Based on the basic information provided by \cite{2016AcA....66..421P}, the color index $(V-I)_0$ helps to estimate the temperatures of these massive binaries, and the temperatures ($T_1$) were derived from the online table \footnote{http$://$www.pas.rochester.edu/$\sim$emamajek/EEM$\_$dwarf$\_$UBVIJHK$\_$colors$\_$Teff.txt} by Mamajek (based mainly on table 5 from \citealt{2013ApJS..208....9P}), their values are shown in Table \ref{table:1}. These binaries belong to the early B-type spectrum, and the effective surface temperatures of star 1 were determined. Consequently, the bolometric albedos $A_1$ = $A_2$ = 1.0 \citep{1969AcA....19..245R} and the gravity-darkening
coefficients $g_1$ = $g_2$ = 1.0 \citep{1967ZA.....65...89L} were adopted. In W-D, the bandpass limb-darkening coefficients were determined according to the logarithmic function. While it was set as free parameters in PHOEBE, and the model of stellar atmospheres was utilized from \cite{2004A&A...419..725C}. Additionally, the mass ratio q ($M_{2}$/$M_{1}$), the orbital inclination $i$, the effective surface temperature of star 2 (T$_{2}$), the equivalent volume radii $r_1$ or  $r_2$ (relative to semimajor axis), are set as free parameters in PHOEBE, the MCMC approach was used by the Emcee package, the detailed settings included 30 walkers and 2000 iterations for each walker. The distribution of these parameters for these four massive binaries is shown in Figure \ref{figure:pb}. The process of determining physical parameters in W-D is similar to these references \citep{2022AJ....164..202L,2023RAA....23a5003W,2023NewA..10202038Z}. Confirmation of the geometric structures and some basic physical parameters for these targets, in particular the PHOEBE solution, can help to determine other parameter values. Then, the third section suggests that three of the four binaries have a third body. In the final step, we tried to include the third light and set the mass ratio as a free parameter, which allowed us to perform the final modeling solution of the light curve.

\begin{figure}
\begin{center}
\includegraphics[width=60mm,height=60mm]{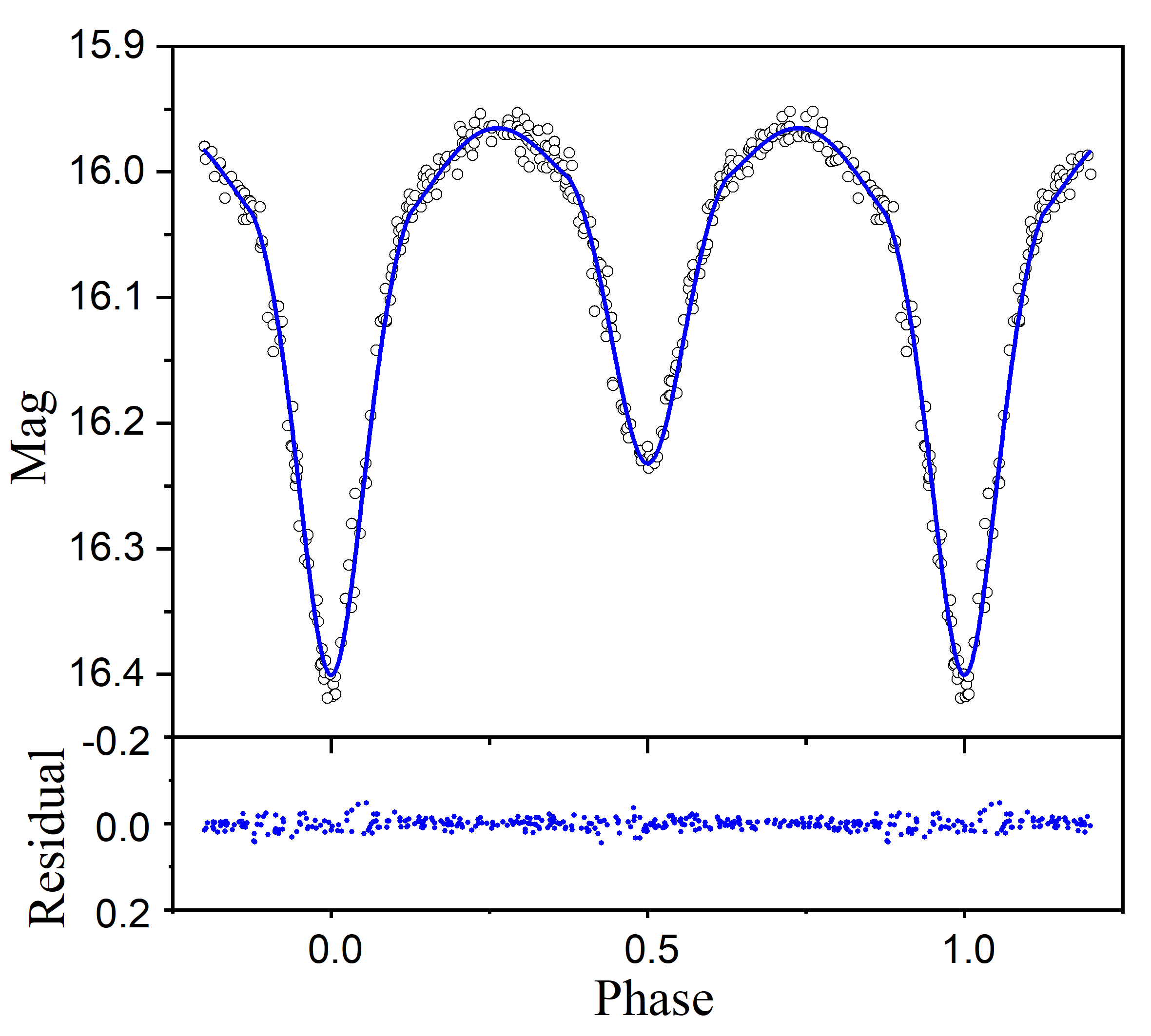}
\includegraphics[width=60mm,height=60mm]{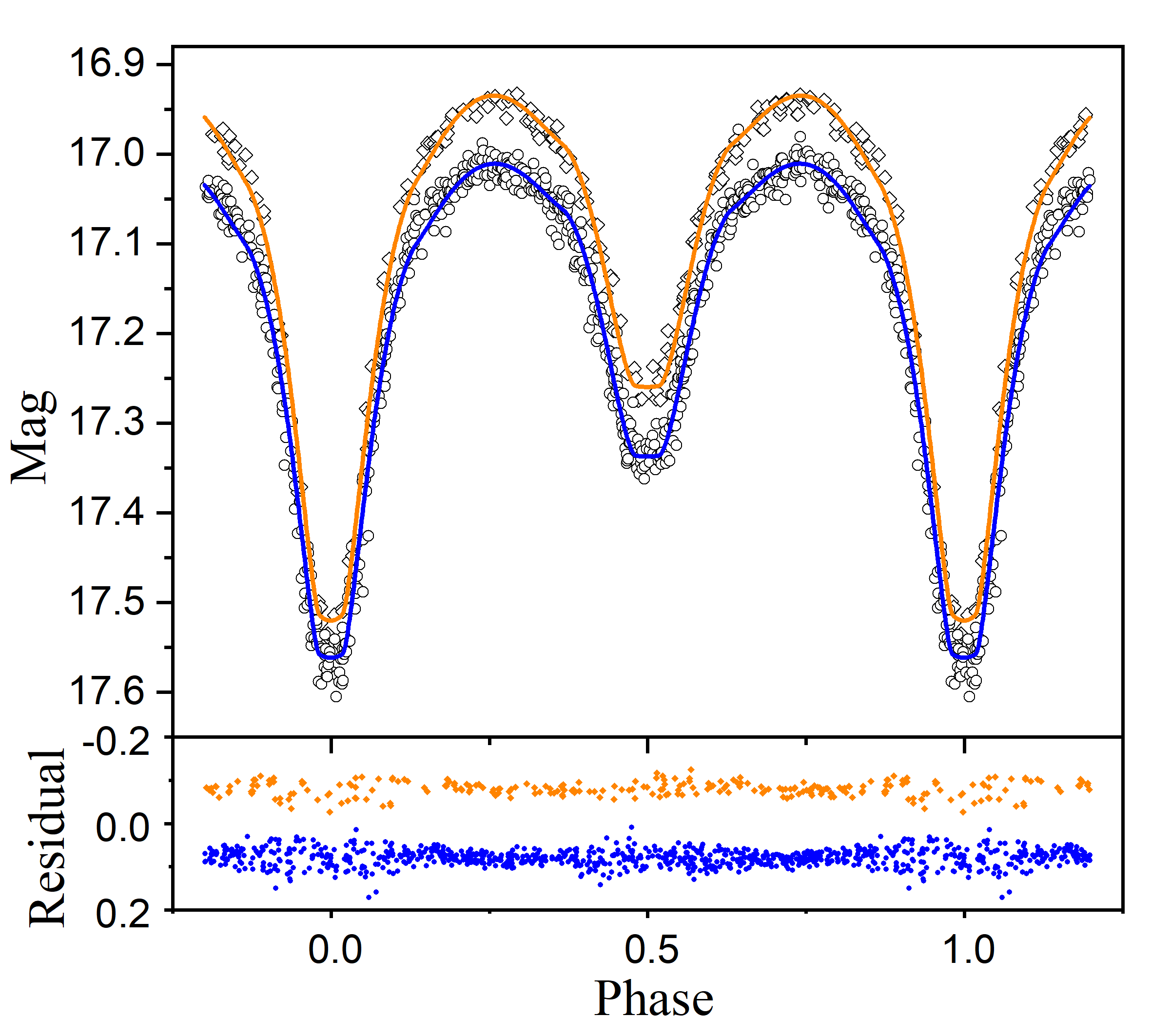}\\
S03065 \hspace{16em} S03065\\
\includegraphics[width=60mm,height=60mm]{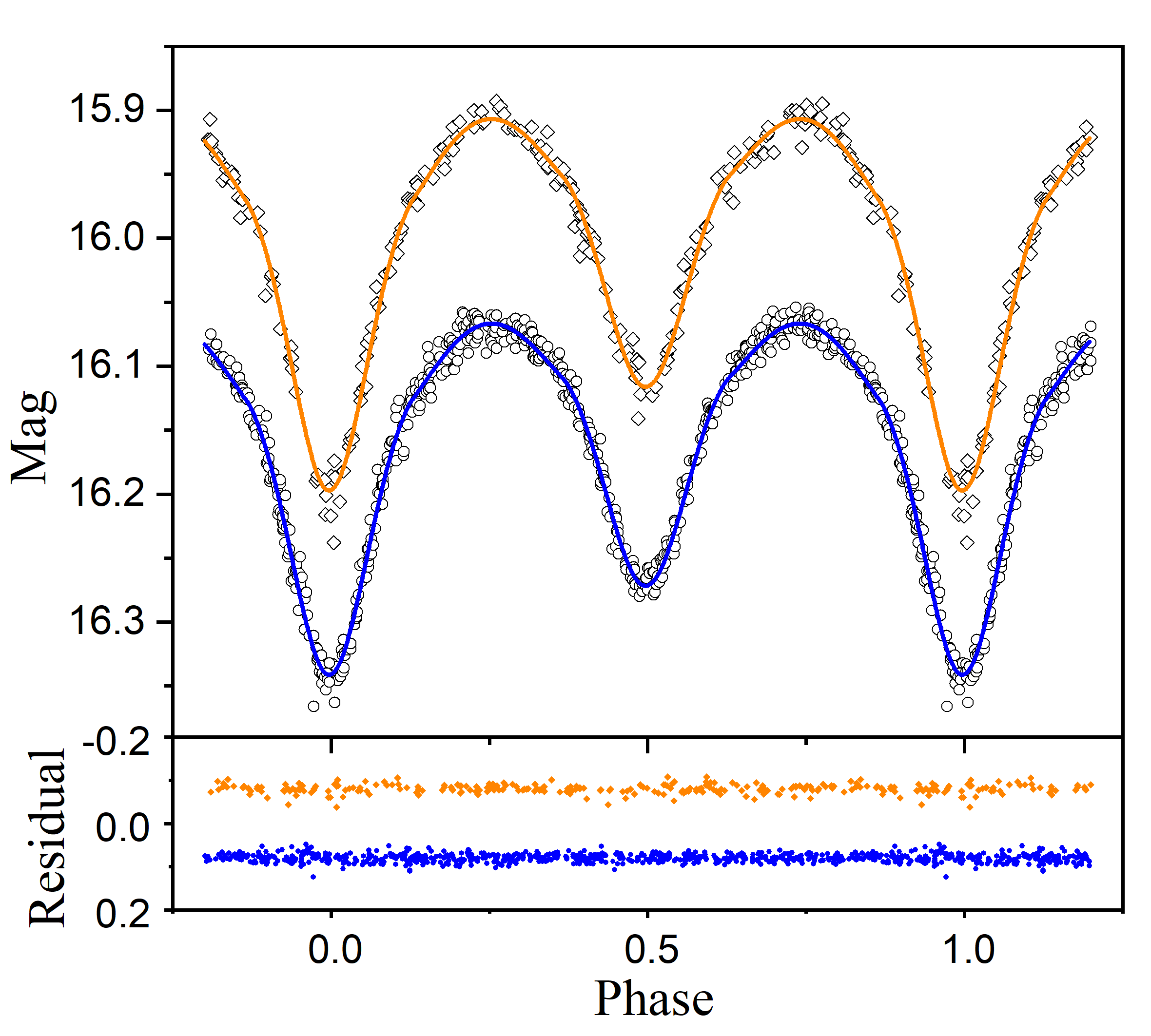}
\includegraphics[width=60mm,height=60mm]{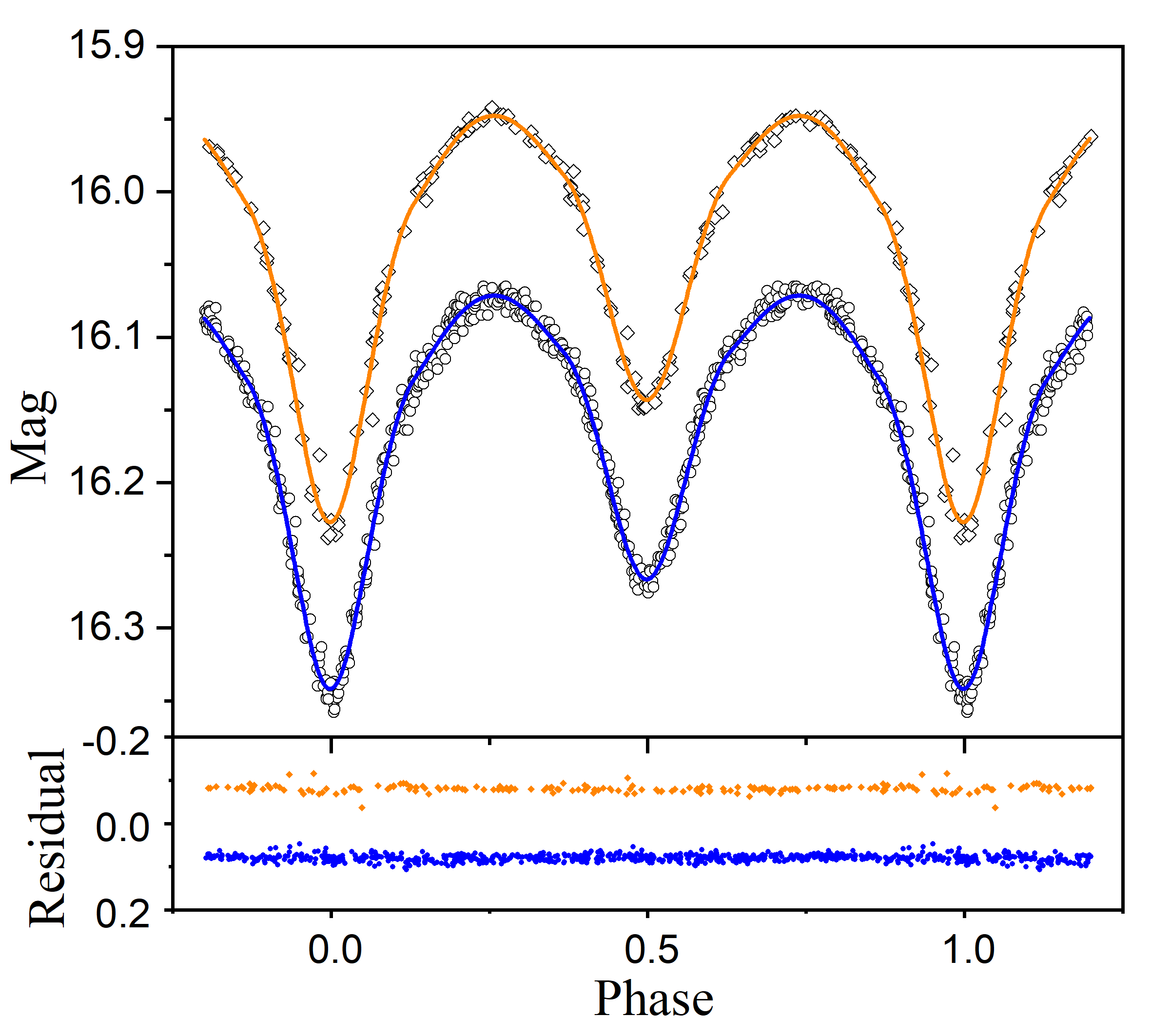}\\
S12631 \hspace{16em} S16873\\
\caption{The solid lines (the orange and blue ones refer to the V band and I band, respectively) and black points represent the theoretical and observational light curves for four semidetached binaries in the LMC; Symbols in the lower subpanels in each panel represent the corresponding residual.}
\end{center}
\label{figure:lc}
\end{figure}

The results of the light curve modeling suggest that these four massive binaries belong to semidetached binaries. Except for S07798, which is the more massive component filling its Roche lobe, the others are the less ones filling their Roche lobes. This distinction is crucial for studying mass transfer in massive binaries, particularly S07798, which may require more attention. The analyses of these light curves are shown in Figure \ref{figure:lc}. The physical parameters of these four massive binaries are obtained as shown in Table \ref{table:timescale}. Where for each massive binary system, there are two sets of solutions available, one that includes the third light and one that does not. $f_{1,2}$ and $\Omega_{1,2}$ are the fill ratios and the Roche potentials of star 1 (S03065, S12631, S16873) or star 2 (S07798), respectively, these components that do not fill their Roche lobes, the value of $f_{1,2}$ is the ratio of the corresponding star volume to the Roche lobe volume.

\section{DISCUSSION AND CONCLUSIONS} \label{subsec:conclusion}

Through created O-C analysis from the minima and derived solution from the light curves in the LMC, this study demonstrates four massive semidetached binaries with a long-term decrease in orbital period. Compared to the number of the massive non-contact binaries (1178) or the massive binaries with EB-type light curves (165), The fraction of massive semidetached binaries showing long-term orbital period decrease is notably small. This rarity adds to their appeal as objects of reliable observational interest. Furthermore, O-C results indicate that three out of the four systems have a third body (75\%), forming triple or higher-order multiple systems. This finding is consistent with the results from \cite{2023ApJ...956...49L} on the proportion of massive binaries with third bodies in the LMC, which are also notably high. \cite{2017ApJS..230...15M} and \cite{2023ASPC..534..275O} studies reveal the fraction of the triple systems fraction is about 35\% in this mass range, suggesting strong observational results for a high incidence of third bodies in massive binaries. This underscores the possibility that the third-body fraction in the LMC might be even higher, potentially influenced by the metallicity of the LMC. It should be noted that the findings may also be influenced by selection effects.

Next, we carried out the third bodies of these massive semidetached binaries to explore the structure of triple systems. The estimates for the third bodies were calculated using Equation 2.
\begin{equation}
{f}(m) = \frac{(M_{3}sini)^3}{(M_{1}+M_{2}+M_{3})^2} = \frac{4\pi^2}{GP_{3}^2}\times(a_{12}sini)^3
\end{equation}
where f(m) is the mass function of the system, $M_{1}$ is shown in Table \ref{table:time}, which is estimated via the color index $(V-I)_0$ by the online table (see also the footnote 5) from Mamajek. The uncertainty is estimated to be 20\% of the mass of the primary component.
The minimum masses of the third bodies ($M_{3}$) were determined to be range between 1.19 and 2.94 $M_{\odot}$. Consequently, the minimum mass of the third body of S16873 is larger than that of its secondary component ($M_2$), which is a B-type spectral companion. This result is consistent with the findings of V357 Cas\citep{2022MNRAS.514.1206L} and ZZ Cas\citep{2022PASJ...74..533L} from the Milky Way, but the orbital periods of these two binaries are long-term increases, and the third body is much farther away from the central binary, which is very intriguing. In the triple systems S03065 and S07798, the mass of the third body is equal to $M_2$ only when its orbital inclination is very low. This may imply that these two massive binaries are accompanied by low-mass third bodies, This is similar to the distribution of third bodies in massive binary systems from \cite{2023ApJ...956...49L}. In these massive binaries with a third body, the results of light curve modeling may also reveal the presence of the third light from the third body. Therefore, these massive binaries and their third components may be forming simultaneously, consistent with the filament fragmentation theory.\citep{2017ApJS..230...15M,2018ApJ...854...44M,2020MNRAS.491.5158T,2023ASPC..534..275O}.

Based on the criterion of dynamical stability for the triple system \citep{2001MNRAS.321..398M,2018ApJ...854...44M,2023A&A...678A..60K}, they provided the following formula: 
\begin{equation}
\left( \frac{a_{\text{out},0}}{a_{\text{in},0}} \right)_{\text{crit}} = \frac{2.8}{1 - e_{\text{out},0}} \left( 1 - 0.3 \frac{i_{\text{tot},0}}{180^\circ} \right) \left[ \frac{(1 + q_{\text{out}})(1 + e_{\text{out},0})}{\sqrt{1 - e_{\text{out},0}}} \right]^{\frac{2}{5}}
\end{equation}
where $e_{\text{out},0}$ represents the eccentricity of the third body, $q_{\text{out}}=M_3/(M_1+M_2)$ , and $i_{\text{tot},0}$ is the orbital inclination between the orbits of the inner binary and the outer third body. When $i_{\text{tot},0}=0$ degrees, based on the minimum mass of the third body in these binaries, the values of $\left( \frac{a_{\text{out},0}}{a_{\text{in},0}} \right)_{\text{crit}}$ for the S03065, S07798, and S16873 systems are 15.1, 3.0, and 7.0, respectively. Whereas the actual $\frac{a_{\text{out}}}{a_{\text{in}}}$ values for these three systems are 38.5, 32.1, and 70.8, respectively. They all belong to hierarchical triple-star systems. Based on this criterion, $\frac{a_{\text{out}}}{a_{\text{in}}}>\left( \frac{a_{\text{out}}}{a_{\text{in}}} \right)_{\text{crit}}$, the actual $\frac{a_{\text{out}}}{a_{\text{in}}}$ values for these three systems are all larger than $\left( \frac{a_{\text{out}}}{a_{\text{in}}} \right)_{\text{crit}}$, indicating that the dynamics of these systems are stable.

We identified four semidetached massive binaries undergoing long-term orbital shrinkage in the LMC. Their rates of orbital variation are on the order of $10^{-7} {d \,yr^{-1}}$, with periods around one day. However, only the light curve of S07798 supports Mode 4 interpretation, indicating that its more massive component fills its Roche lobe, transferring mass to the smaller companion, leading to a long-term decrease in orbital period\citep{2022A&A...659A..98S,2022ApJ...924...30L}. In comparison to the thermal timescale, this binary may no longer be in the rapid mass-transfer phase in Case A. Conversely, for S03065, S12631, and S16873, their light curve interpretations suggest these semidetached massive binaries have undergone the stage of mass ratio inversion, the smaller components fill the Roche lobe, transferring mass to their companions, which would normally cause orbital expansion\citep{2013ApJS..207...22Q,2021AJ....162...13L}, contradicting the results exactly. Consequently, which may confirm that these binaries have a significant angular momentum loss, the role it plays is greater than the influence of mass transfer. S03065 and S16873 exhibit the third bodies, likely responsible for angular momentum extraction by the eccentric Kozai-Lidov mechanism  \citep{2014ApJ...793..137N,2016ARA&A..54..441N}. Meanwhile, no third body has been detected for S12631, suggesting angular momentum loss may be due to stellar winds or disk winds, gravitational torques, as performed by \cite{2012ASPC..465..451K} for the angular momentum loss mechanism, and it should also consider the fact that the metallicity of the LMC is lower\citep{2005A&A...443..643Y}. Additionally, the mass transfer timescales of the more massive ones were calculated in these binaries by equations 3\citep{1975MNRAS.170..633P},
\begin{equation}
\frac{dM_1}{dt} = \frac{M_1 M_2}{3P(M_1 - M_2)} \times \frac{dP}{dt},
\end{equation}
the corresponding thermal timescale $\tau_\textrm{th}$ derived from the following equation 4\citep{1971ARA&A...9..183P},
\begin{equation}
    \tau_\textrm{th} = \frac{GM_{1,2}^2}{R_{1,2} L_{1,2}},
\end{equation}
and the luminosity of these massive was calculated by the formula as follows,
\begin{equation}
\frac{L_{1,2}}{L_{\odot}} = (\frac{R_{1,2}}{R_{\odot}})^2\times(\frac{T_{1,2}}{T_{\odot}})^4.
\end{equation}
The values are listed in the table \ref{table:time}. The mass-transfer timescales greatly exceed the thermal timescales, both of which relate to the component filling its Roche lobe for these four massive binaries. This evidence may demonstrate that the angular momentum loss dominates the evolution of these massive binaries. However, it should be noted that these findings consider only the currently observed data. Further observations are necessary to explore them.

\begin{figure}
\begin{center}
\includegraphics[angle=0,scale=0.5]{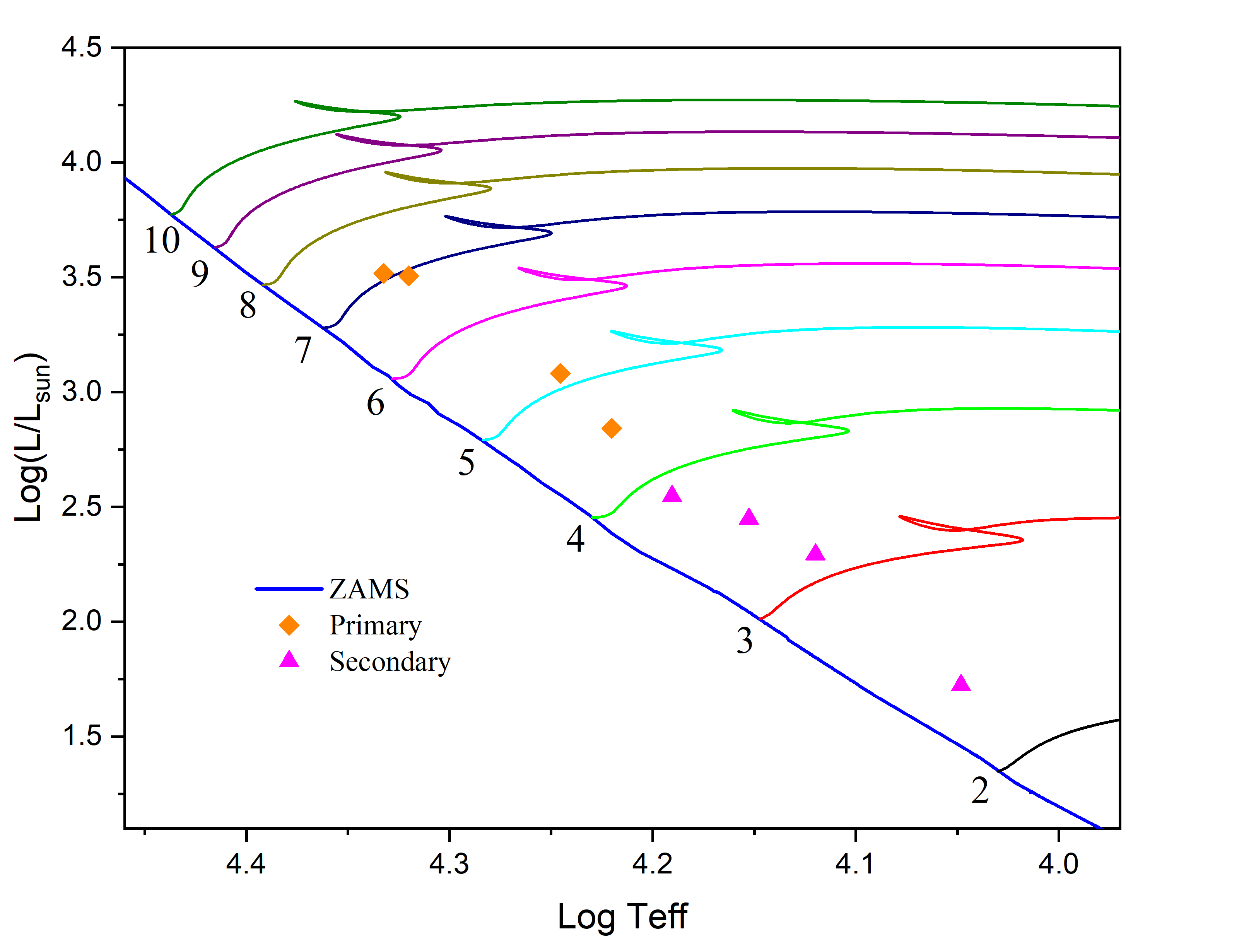}
\caption{The stellar evolutionary models of four semidetached massive binaries in the LMC. The blue solid line is the ZAMS. The other solid lines represent the evolutionary tracks for the different masses (2-10$M_\odot$). }
\end{center}
\label{figure:ev}
\end{figure}

After that, we try to understand the state of evolution, and the stellar evolutionary tracks and isochrones with a mean metallicity of $[Fe/{H}] = -0.42$ in the LMC are created based on the CMD 3.7 edition \citep{2012MNRAS.427..127B,2014MNRAS.444.2525C,2015MNRAS.452.1068C,2014MNRAS.445.4287T,2017ApJ...835...77M,2019MNRAS.485.5666P}, the Hertzsprung-Russell diagram is shown in Figure \ref{figure:ev}. The results indicate that both the primary and secondary components of these semidetached massive binaries are located in the main sequence.

In summary, there are several aspects of this work are worth consolidating. Firstly, many studies suggest that the proportion of third bodies in massive binaries is very high. These observational results help to confirm this. Importantly, the findings imply that third bodies may play a crucial role in the loss of angular momentum, thereby dominating the influence on the evolution of massive binaries. Secondly, these attractive massive binaries, which have a semidetached geometric configuration with long-term orbital decrease, discover the physical phenomena resulting from the combined effects of mass transfer theories and angular momentum loss mechanisms. It serves as an ideal laboratory to study the mechanism of angular momentum loss in massive systems. These results provide a reasonable explanation for the formation and evolution of massive binary stars.

\acknowledgments
This work is partly supported by International Cooperation Projects of the National Key R\&D Program (No. 2022YFE0127300) and the Chinese Natural Science Foundation (Grant Nos. 12303040, 11933008, 11873017). It is also partly supported by the basic research project of Yunnan Province (Grant Nos. 202401AT070143) and Yunnan Revitalization Talent Support Program. This paper makes use of publicly available data from the OGLE (https://http://ogle.astrouw.edu.pl/) and the EROS-2 survey in the VizieR archives (https://vizier.cds.unistra.fr/viz-bin/VizieR). All of the TESS data used in this paper can be found through the MAST: \dataset[doi:10.17909/r1b8-aj60]{https://archive.stsci.edu/doi/resolve/resolve.html?doi=10.17909/r1b8-aj60}. The authors thank the OGLE, EROS-2, TESS teams for providing their observation data, which creates an opportunity for us to accomplish this work.

\appendix

\section{eclipsing times table}
The eclipse times of the four semidetached massive binaries are shown in Table \ref{tab:min}.

\begin{longtable}{llllllllccccc}
\caption{The eclipsing times of four semidetached massive binaries.}\\
\hline\hline
Name &Eclipse-times    &Errors  &p/s   &Source     &Eclipse-times   &Errors  &p/s &Source  \\
 &HJD$-$2\,450\,000      &$\pm$days  &        &     &HJD$-$2\,450\,000      &$\pm$days &    & \\
\hline
\endfirsthead 

\caption{(Continued)}\\
\hline\hline
Name &Eclipse-times    &Errors  &p/s   &Source     &Eclipse-times   &Errors  &p/s &Source  \\
 &HJD$-$2\,450\,000      &$\pm$days  &        &     &HJD$-$2\,450\,000      &$\pm$days &    & \\
\hline \endhead 

\hline
\multicolumn{6}{r}{\textsl{(Continued)}}\\
\endfoot

\endlastfoot
S03065	&	2507.79499 	&	0.00383 	&	p	&	OGLE III	&	9245.41831 	&	0.00507 	&	s	&	TESS	\\
	&	2600.14850 	&	0.00522 	&	s	&	OGLE III	&	9245.95394 	&	0.00546 	&	p	&	TESS	\\
	&	2815.10209 	&	0.00560 	&	s	&	OGLE III	&	9247.58366 	&	0.00704 	&	s	&	TESS	\\
	&	2964.70343 	&	0.00342 	&	p	&	OGLE III	&	9248.11730 	&	0.00494 	&	p	&	TESS	\\
	&	3112.68799 	&	0.00296 	&	p	&	OGLE III	&	9249.74722 	&	0.00674 	&	s	&	TESS	\\
	&	3163.99461 	&	0.00310 	&	s	&	OGLE III	&	9250.27851 	&	0.00493 	&	p	&	TESS	\\
	&	3456.73026 	&	0.00451 	&	s	&	OGLE III	&	9251.90436 	&	0.00765 	&	s	&	TESS	\\
	&	3458.35156 	&	0.00304 	&	p	&	OGLE III	&	9252.43981 	&	0.00559 	&	p	&	TESS	\\
	&	3795.37504 	&	0.00188 	&	p	&	OGLE III	&	9253.52660 	&	0.00837 	&	p	&	TESS	\\
	&	3844.52674 	&	0.00365 	&	s	&	OGLE III	&	9311.84787 	&	0.00560 	&	p	&	TESS	\\
	&	4056.78091 	&	0.00296 	&	p	&	OGLE III	&	9312.92819 	&	0.00569 	&	p	&	TESS	\\
	&	4071.36473 	&	0.00237 	&	s	&	OGLE III	&	9315.08771 	&	0.00685 	&	p	&	TESS	\\
	&	4447.26947 	&	0.00262 	&	s	&	OGLE III	&	9323.72454 	&	0.00519 	&	p	&	TESS	\\
	&	4548.26660 	&	0.00203 	&	p	&	OGLE III	&	9324.81564 	&	0.00553 	&	p	&	TESS	\\
	&	4696.79210 	&	0.00203 	&	s	&	OGLE III	&	9326.97383 	&	0.00527 	&	p	&	TESS	\\
	&	4809.67027 	&	0.00199 	&	p	&	OGLE III	&	9329.13317 	&	0.00612 	&	p	&	TESS	\\
	&	5398.91006 	&	0.00206 	&	s	&	OGLE IIV	&	9331.29451 	&	0.00682 	&	p	&	TESS	\\
	&	5432.93345 	&	0.00172 	&	p	&	OGLE IIV	&	9332.36840 	&	0.00924 	&	p	&	TESS	\\
	&	5583.07835 	&	0.00104 	&	p	&	OGLE IIV	&	10043.11924 	&	0.00503 	&	p	&	TESS	\\
	&	5607.37943 	&	0.00284 	&	s	&	OGLE IIV	&	10043.65449 	&	0.00840 	&	s	&	TESS	\\
	&	5712.69817 	&	0.00149 	&	p	&	OGLE IIV	&	10045.27776 	&	0.00480 	&	p	&	TESS	\\
	&	5758.60371 	&	0.00182 	&	s	&	OGLE IIV	&	10045.81637 	&	0.00763 	&	s	&	TESS	\\
	&	5911.99064 	&	0.00165 	&	s	&	OGLE IIV	&	10047.97649 	&	0.00873 	&	s	&	TESS	\\
	&	5927.65309 	&	0.00467 	&	p	&	OGLE IIV	&	10048.51433 	&	0.00510 	&	p	&	TESS	\\
	&	6042.14733 	&	0.00179 	&	p	&	OGLE IIV	&	10050.13664 	&	0.00952 	&	s	&	TESS	\\
	&	6086.97757 	&	0.00218 	&	s	&	OGLE IIV	&	10050.67595 	&	0.00531 	&	p	&	TESS	\\
	&	6376.99816 	&	0.00504 	&	p	&	OGLE IIV	&	10052.83867 	&	0.00533 	&	p	&	TESS	\\
	&	6534.70136 	&	0.00538 	&	p	&	OGLE IIV	&	10053.37111 	&	0.00995 	&	s	&	TESS	\\
	&	8330.51424 	&	0.00768 	&	s	&	TESS	&	10054.99723 	&	0.00532 	&	p	&	TESS	\\
	&	8331.05532 	&	0.00561 	&	p	&	TESS	&	10057.15578 	&	0.00488 	&	p	&	TESS	\\
	&	8336.98989 	&	0.00845 	&	s	&	TESS	&	10057.69116 	&	0.00744 	&	s	&	TESS	\\
	&	8337.53746 	&	0.00746 	&	p	&	TESS	&	10059.86215 	&	0.00732 	&	s	&	TESS	\\
	&	8343.47046 	&	0.00713 	&	s	&	TESS	&	10060.39806 	&	0.00435 	&	p	&	TESS	\\
	&	8344.01306 	&	0.00956 	&	p	&	TESS	&	10062.02388 	&	0.00684 	&	s	&	TESS	\\
	&	8349.41169 	&	0.01357 	&	p	&	TESS	&	10062.56092 	&	0.00436 	&	p	&	TESS	\\
	&	8428.80680 	&	0.00580 	&	s	&	TESS	&	10064.17798 	&	0.00692 	&	s	&	TESS	\\
	&	8429.34767 	&	0.00453 	&	p	&	TESS	&	10064.71883 	&	0.00463 	&	p	&	TESS	\\
	&	8432.58536 	&	0.00606 	&	p	&	TESS	&	10066.33777 	&	0.00961 	&	s	&	TESS	\\
	&	8434.20367 	&	0.00599 	&	s	&	TESS	&	10066.87995 	&	0.00647 	&	p	&	TESS	\\
	&	8575.71002 	&	0.01038 	&	s	&	TESS	&	10070.65577 	&	0.00847 	&	s	&	TESS	\\
	&	8576.25400 	&	0.00503 	&	p	&	TESS	&	10071.20269 	&	0.00676 	&	p	&	TESS	\\
	&	8582.72842 	&	0.00648 	&	p	&	TESS	&	10073.36760 	&	0.00578 	&	p	&	TESS	\\
	&	8583.27917 	&	0.00628 	&	s	&	TESS	&	10075.52667 	&	0.00625 	&	p	&	TESS	\\
	&	8590.29283 	&	0.00510 	&	p	&	TESS	&	10077.68580 	&	0.00667 	&	p	&	TESS	\\
	&	8590.83844 	&	0.00707 	&	s	&	TESS	&	10078.21862 	&	0.00942 	&	s	&	TESS	\\
	&	8593.52561 	&	0.00662 	&	p	&	TESS	&	10079.83662 	&	0.00643 	&	p	&	TESS	\\
	&	8634.03680 	&	0.00656 	&	s	&	TESS	&	10080.37696 	&	0.00843 	&	s	&	TESS	\\
	&	8634.58039 	&	0.00356 	&	p	&	TESS	&	10084.16143 	&	0.00627 	&	p	&	TESS	\\
	&	8641.60296 	&	0.00725 	&	s	&	TESS	&	10087.40447 	&	0.00506 	&	p	&	TESS	\\
	&	8642.14544 	&	0.00361 	&	p	&	TESS	&	10090.63964 	&	0.00612 	&	p	&	TESS	\\
	&	8648.08592 	&	0.00339 	&	s	&	TESS	&	10093.34832 	&	0.00843 	&	s	&	TESS	\\
	&	8648.62102 	&	0.00342 	&	p	&	TESS	&	10093.87846 	&	0.00779 	&	p	&	TESS	\\
	&	8651.32641 	&	0.00549 	&	s	&	TESS	&	10094.96373 	&	0.00999 	&	p	&	TESS	\\
	&	8651.86008 	&	0.00577 	&	p	&	TESS									\\\hline
S07798	&	585.45299 	&	0.00249 	&	p	&	OGLE II	&	5333.94002 	&	0.00169 	&	s	&	OGLE IV	\\
	&	604.26242 	&	0.00150 	&	s	&	OGLE II	&	5359.94984 	&	0.00267 	&	p	&	OGLE IV	\\
	&	769.14911 	&	0.00208 	&	s	&	OGLE II	&	5539.64649 	&	0.00248 	&	s	&	OGLE IV	\\
	&	783.15711 	&	0.00216 	&	p	&	OGLE II	&	5540.04519 	&	0.00090 	&	p	&	OGLE IV	\\
	&	1034.08934 	&	0.00204 	&	s	&	OGLE II	&	5588.87117 	&	0.00132 	&	p	&	OGLE IV	\\
	&	1072.10897 	&	0.00175 	&	p	&	OGLE II	&	5610.08197 	&	0.00280 	&	s	&	OGLE IV	\\
	&	1344.65388 	&	0.00205 	&	s	&	OGLE II	&	5646.10051 	&	0.00315 	&	s	&	OGLE IV	\\
	&	1347.45720 	&	0.00279 	&	p	&	OGLE II	&	5648.90084 	&	0.00168 	&	p	&	OGLE IV	\\
	&	1479.12669 	&	0.00405 	&	s	&	OGLE II	&	5740.54897 	&	0.00296 	&	s	&	OGLE IV	\\
	&	1542.76607 	&	0.00309 	&	p	&	OGLE II	&	5755.35623 	&	0.00133 	&	p	&	OGLE IV	\\
	&	1585.19047 	&	0.00305 	&	p	&	OGLE II	&	5857.00945 	&	0.00167 	&	p	&	OGLE IV	\\
	&	2628.52646 	&	0.00273 	&	s	&	OGLE III	&	5875.01753 	&	0.00267 	&	s	&	OGLE IV	\\
	&	2705.76573 	&	0.00173 	&	p	&	OGLE III	&	5926.64424 	&	0.00227 	&	p	&	OGLE IV	\\
	&	2850.24242 	&	0.00351 	&	s	&	OGLE III	&	5939.85131 	&	0.00194 	&	s	&	OGLE IV	\\
	&	2872.25049 	&	0.00145 	&	p	&	OGLE III	&	5977.47097 	&	0.00189 	&	s	&	OGLE IV	\\
	&	3107.97701 	&	0.00480 	&	s	&	OGLE III	&	5989.07450 	&	0.00174 	&	p	&	OGLE IV	\\
	&	3173.20614 	&	0.00198 	&	p	&	OGLE III	&	6049.50726 	&	0.00242 	&	s	&	OGLE IV	\\
	&	3421.33523 	&	0.00193 	&	p	&	OGLE III	&	6077.92095 	&	0.00312 	&	p	&	OGLE IV	\\
	&	3444.15085 	&	0.00271 	&	s	&	OGLE III	&	6141.15416 	&	0.00251 	&	p	&	OGLE IV	\\
	&	3786.32851 	&	0.00178 	&	p	&	OGLE III	&	6160.76354 	&	0.00196 	&	s	&	OGLE IV	\\
	&	3812.34486 	&	0.00233 	&	s	&	OGLE III	&	6242.80669 	&	0.00162 	&	p	&	OGLE IV	\\
	&	4136.11440 	&	0.00129 	&	p	&	OGLE III	&	6248.80806 	&	0.00295 	&	s	&	OGLE IV	\\
	&	4201.35216 	&	0.00221 	&	s	&	OGLE III	&	6311.24116 	&	0.00229 	&	s	&	OGLE IV	\\
	&	4370.64055 	&	0.00286 	&	p	&	OGLE III	&	6312.44340 	&	0.00232 	&	p	&	OGLE IV	\\
	&	4459.89142 	&	0.00354 	&	s	&	OGLE III	&	6439.30937 	&	0.00214 	&	s	&	OGLE IV	\\
	&	4729.63653 	&	0.00319 	&	s	&	OGLE III	&	6510.14935 	&	0.00269 	&	p	&	OGLE IV	\\
	&	4819.68079 	&	0.00280 	&	p	&	OGLE III	&	6562.17646 	&	0.00261 	&	p	&	OGLE IV	\\
	&	5103.42769 	&	0.00458 	&	s	&	OGLE IV	&	6592.19378 	&	0.00277 	&	s	&	OGLE IV	\\
	&	5119.83409 	&	0.00374 	&	p	&	OGLE IV									\\\hline
S12631	&	658.34198 	&	0.00510 	&	p	&	EROS-2	&	8577.88698 	&	0.00603 	&	p	&	TESS	\\
	&	877.20553 	&	0.00574 	&	s	&	EROS-2	&	8578.41635 	&	0.00608 	&	s	&	TESS	\\
	&	1054.86295 	&	0.00443 	&	s	&	EROS-2	&	8581.09735 	&	0.00676 	&	p	&	TESS	\\
	&	1075.72857 	&	0.00455 	&	p	&	EROS-2	&	8581.63369 	&	0.00849 	&	s	&	TESS	\\
	&	1189.70975 	&	0.00529 	&	s	&	EROS-2	&	8583.77395 	&	0.00808 	&	s	&	TESS	\\
	&	1198.80488 	&	0.00350 	&	p	&	EROS-2	&	8584.30733 	&	0.00689 	&	p	&	TESS	\\
	&	1331.51280 	&	0.00482 	&	p	&	EROS-2	&	8586.98485 	&	0.00781 	&	s	&	TESS	\\
	&	1342.74876 	&	0.00828 	&	s	&	EROS-2	&	8587.51755 	&	0.00563 	&	p	&	TESS	\\
	&	1481.34307 	&	0.00548 	&	p	&	EROS-2	&	8589.11864 	&	0.00603 	&	s	&	TESS	\\
	&	1506.49372 	&	0.01404 	&	s	&	EROS-2	&	8589.65822 	&	0.00449 	&	p	&	TESS	\\
	&	1627.43347 	&	0.01596 	&	s	&	EROS-2	&	8591.79974 	&	0.00556 	&	p	&	TESS	\\
	&	1678.26207 	&	0.00480 	&	p	&	EROS-2	&	8592.33319 	&	0.00587 	&	s	&	TESS	\\
	&	1824.88178 	&	0.00400 	&	p	&	EROS-2	&	8592.87054 	&	0.00610 	&	p	&	TESS	\\
	&	1925.47649 	&	0.00509 	&	p	&	EROS-2	&	8593.40187 	&	0.00602 	&	s	&	TESS	\\
	&	1946.34993 	&	0.02050 	&	s	&	EROS-2	&	8593.93860 	&	0.00826 	&	p	&	TESS	\\
	&	2113.83446 	&	0.00542 	&	p	&	EROS-2	&	8594.47655 	&	0.00782 	&	s	&	TESS	\\
	&	2149.69247 	&	0.01200 	&	s	&	EROS-2	&	9070.17560 	&	0.00809 	&	p	&	TESS	\\
	&	2252.96627 	&	0.01150 	&	p	&	EROS-2	&	9070.70768 	&	0.00831 	&	s	&	TESS	\\
	&	2574.56642 	&	0.00306 	&	s	&	OGLE III	&	9073.39134 	&	0.01026 	&	p	&	TESS	\\
	&	2610.41754 	&	0.00256 	&	p	&	OGLE III	&	9075.52883 	&	0.00797 	&	p	&	TESS	\\
	&	2873.15458 	&	0.00323 	&	s	&	OGLE III	&	9077.66916 	&	0.00701 	&	p	&	TESS	\\
	&	2902.58538 	&	0.00213 	&	p	&	OGLE III	&	9078.20978 	&	0.00919 	&	s	&	TESS	\\
	&	3187.26180 	&	0.00239 	&	p	&	OGLE III	&	9079.81730 	&	0.00760 	&	p	&	TESS	\\
	&	3220.97187 	&	0.00410 	&	s	&	OGLE III	&	9080.34467 	&	0.00836 	&	s	&	TESS	\\
	&	3527.58708 	&	0.00228 	&	p	&	OGLE III	&	9081.96061 	&	0.00994 	&	p	&	TESS	\\
	&	3533.47147 	&	0.00477 	&	s	&	OGLE III	&	9082.47726 	&	0.00878 	&	s	&	TESS	\\
	&	3923.56437 	&	0.00273 	&	p	&	OGLE III	&	9084.09898 	&	0.00791 	&	p	&	TESS	\\
	&	3967.97790 	&	0.00399 	&	s	&	OGLE III	&	9084.62157 	&	0.00757 	&	s	&	TESS	\\
	&	4196.46811 	&	0.00309 	&	p	&	OGLE III	&	9146.15994 	&	0.00702 	&	p	&	TESS	\\
	&	4350.03964 	&	0.00531 	&	s	&	OGLE III	&	9148.29602 	&	0.00786 	&	p	&	TESS	\\
	&	4658.26256 	&	0.00499 	&	s	&	OGLE III	&	9148.84578 	&	0.00787 	&	s	&	TESS	\\
	&	4737.99600 	&	0.00296 	&	p	&	OGLE III	&	9150.44533 	&	0.00789 	&	p	&	TESS	\\
	&	4965.41262 	&	0.00569 	&	s	&	OGLE III	&	9150.97990 	&	0.00692 	&	s	&	TESS	\\
	&	5271.49436 	&	0.00763 	&	s	&	OGLE IIV	&	9152.58684 	&	0.00771 	&	p	&	TESS	\\
	&	5338.38083 	&	0.00231 	&	p	&	OGLE IIV	&	9153.12255 	&	0.00756 	&	s	&	TESS	\\
	&	5493.56046 	&	0.00215 	&	p	&	OGLE IIV	&	9154.18863 	&	0.00920 	&	s	&	TESS	\\
	&	5512.28762 	&	0.00442 	&	s	&	OGLE IIV	&	9154.72993 	&	0.01179 	&	p	&	TESS	\\
	&	5569.54383 	&	0.00375 	&	p	&	OGLE IIV	&	9157.38668 	&	0.01163 	&	s	&	TESS	\\
	&	5574.35842 	&	0.00317 	&	s	&	OGLE IIV	&	9159.53341 	&	0.00808 	&	s	&	TESS	\\
	&	5642.31646 	&	0.00353 	&	p	&	OGLE IIV	&	9160.06284 	&	0.00818 	&	p	&	TESS	\\
	&	5648.20418 	&	0.00261 	&	s	&	OGLE IIV	&	9162.21251 	&	0.00648 	&	p	&	TESS	\\
	&	5722.58210 	&	0.00477 	&	p	&	OGLE IIV	&	9162.73960 	&	0.00964 	&	s	&	TESS	\\
	&	5748.80527 	&	0.00338 	&	s	&	OGLE IIV	&	9164.35979 	&	0.00705 	&	p	&	TESS	\\
	&	5835.49072 	&	0.00232 	&	s	&	OGLE IIV	&	9164.89451 	&	0.00935 	&	s	&	TESS	\\
	&	5848.86718 	&	0.00401 	&	p	&	OGLE IIV	&	9166.48819 	&	0.00952 	&	p	&	TESS	\\
	&	5905.05461 	&	0.00337 	&	s	&	OGLE IIV	&	9167.03036 	&	0.00814 	&	s	&	TESS	\\
	&	5921.64240 	&	0.00247 	&	p	&	OGLE IIV	&	9168.63693 	&	0.00920 	&	p	&	TESS	\\
	&	5958.56553 	&	0.00245 	&	s	&	OGLE IIV	&	10041.91128 	&	0.00937 	&	p	&	TESS	\\
	&	5962.31165 	&	0.00244 	&	p	&	OGLE IIV	&	10041.91243 	&	0.01010 	&	p	&	TESS	\\
	&	6025.98895 	&	0.00530 	&	s	&	OGLE IIV	&	10042.45911 	&	0.00944 	&	s	&	TESS	\\
	&	6030.80493 	&	0.00300 	&	p	&	OGLE IIV	&	10043.52180 	&	0.01083 	&	s	&	TESS	\\
	&	6131.93788 	&	0.00727 	&	s	&	OGLE IIV	&	10045.12151 	&	0.00968 	&	p	&	TESS	\\
	&	6139.97279 	&	0.00299 	&	p	&	OGLE IIV	&	10046.20745 	&	0.01107 	&	p	&	TESS	\\
	&	6302.63339 	&	0.00236 	&	p	&	OGLE IIV	&	10046.20745 	&	0.01107 	&	p	&	TESS	\\
	&	6326.71158 	&	0.00505 	&	s	&	OGLE IIV	&	10050.49410 	&	0.01056 	&	p	&	TESS	\\
	&	6423.56688 	&	0.00292 	&	p	&	OGLE IIV	&	10050.49450 	&	0.01095 	&	p	&	TESS	\\
	&	6450.85466 	&	0.00490 	&	s	&	OGLE IIV	&	10052.62018 	&	0.01156 	&	p	&	TESS	\\
	&	6655.27022 	&	0.01421 	&	s	&	OGLE IIV	&	10053.69800 	&	0.01002 	&	p	&	TESS	\\
	&	8328.53089 	&	0.00716 	&	p	&	TESS	&	10054.21585 	&	0.01102 	&	s	&	TESS	\\
	&	8329.07342 	&	0.00862 	&	s	&	TESS	&	10054.22369 	&	0.01164 	&	s	&	TESS	\\
	&	8330.13909 	&	0.00771 	&	s	&	TESS	&	10054.76256 	&	0.01041 	&	p	&	TESS	\\
	&	8330.67165 	&	0.00548 	&	p	&	TESS	&	10056.89794 	&	0.01000 	&	p	&	TESS	\\
	&	8332.81323 	&	0.00559 	&	p	&	TESS	&	10056.91368 	&	0.01008 	&	p	&	TESS	\\
	&	8333.34611 	&	0.00854 	&	s	&	TESS	&	10098.62952 	&	0.00947 	&	p	&	TESS	\\
	&	8334.95422 	&	0.00590 	&	p	&	TESS	&	10098.63051 	&	0.00972 	&	p	&	TESS	\\
	&	8335.48315 	&	0.00894 	&	s	&	TESS	&	10099.18358 	&	0.01296 	&	s	&	TESS	\\
	&	8336.02746 	&	0.00834 	&	p	&	TESS	&	10099.18449 	&	0.01367 	&	s	&	TESS	\\
	&	8337.09798 	&	0.00846 	&	p	&	TESS	&	10099.71249 	&	0.01231 	&	p	&	TESS	\\
	&	8574.14564 	&	0.00961 	&	s	&	TESS	&	10101.31429 	&	0.01209 	&	s	&	TESS	\\
	&	8574.67891 	&	0.00648 	&	p	&	TESS	&	10101.31462 	&	0.01360 	&	s	&	TESS	\\
	&	8576.27677 	&	0.00617 	&	s	&	TESS	&	10101.84585 	&	0.01393 	&	p	&	TESS	\\
	&	8576.81845 	&	0.00625 	&	p	&	TESS	&	10102.93228 	&	0.00974 	&	p	&	TESS	\\\hline
S16873	&	474.17024 	&	0.00144 	&	p	&	OGLE II	&	5507.34678 	&	0.00167 	&	p	&	OGLE IV	\\
	&	477.94944 	&	0.00369 	&	s	&	OGLE II	&	5571.03077 	&	0.00137 	&	p	&	OGLE IV	\\
	&	489.28068 	&	0.00187 	&	p	&	OGLE II	&	5618.52439 	&	0.00149 	&	p	&	OGLE IV	\\
	&	489.82243 	&	0.00403 	&	s	&	OGLE II	&	5702.17555 	&	0.00251 	&	s	&	OGLE IV	\\
	&	500.61182 	&	0.00415 	&	s	&	OGLE II	&	5711.34996 	&	0.00246 	&	p	&	OGLE IV	\\
	&	511.94722 	&	0.00221 	&	p	&	OGLE II	&	5799.31867 	&	0.00253 	&	s	&	OGLE IV	\\
	&	531.91406 	&	0.00486 	&	s	&	OGLE II	&	5803.09651 	&	0.00154 	&	p	&	OGLE IV	\\
	&	535.69128 	&	0.00376 	&	p	&	OGLE II	&	5881.35189 	&	0.00219 	&	s	&	OGLE IV	\\
	&	645.78434 	&	0.00302 	&	p	&	OGLE II	&	5884.05034 	&	0.00142 	&	p	&	OGLE IV	\\
	&	662.51892 	&	0.00459 	&	s	&	OGLE II	&	5932.62321 	&	0.00176 	&	p	&	OGLE IV	\\
	&	834.13795 	&	0.00483 	&	s	&	OGLE II	&	5934.24327 	&	0.00209 	&	s	&	OGLE IV	\\
	&	872.45468 	&	0.00291 	&	p	&	OGLE II	&	5996.85072 	&	0.00250 	&	s	&	OGLE IV	\\
	&	1094.80831 	&	0.00472 	&	p	&	OGLE II	&	5998.47002 	&	0.00232 	&	p	&	OGLE IV	\\
	&	1098.58543 	&	0.00449 	&	s	&	OGLE II	&	6099.93490 	&	0.00223 	&	p	&	OGLE IV	\\
	&	1347.37843 	&	0.00333 	&	p	&	OGLE II	&	6116.66728 	&	0.00202 	&	s	&	OGLE IV	\\
	&	1349.00346 	&	0.00490 	&	s	&	OGLE II	&	6255.91174 	&	0.00244 	&	s	&	OGLE IV	\\
	&	1507.66680 	&	0.00420 	&	s	&	OGLE II	&	6267.24554 	&	0.00291 	&	p	&	OGLE IV	\\
	&	1528.71112 	&	0.00441 	&	p	&	OGLE II	&	6373.56640 	&	0.00434 	&	s	&	OGLE IV	\\
	&	2449.99211 	&	0.00726 	&	s	&	OGLE III	&	6414.04493 	&	0.00436 	&	p	&	OGLE IV	\\
	&	2529.32637 	&	0.00274 	&	p	&	OGLE III	&	6529.54491 	&	0.00506 	&	p	&	OGLE IV	\\
	&	2628.63109 	&	0.00243 	&	p	&	OGLE III	&	8388.25559 	&	0.00949 	&	p	&	TESS	\\
	&	2632.40846 	&	0.00392 	&	s	&	OGLE III	&	8388.79562 	&	0.01223 	&	s	&	TESS	\\
	&	2646.98176 	&	0.00265 	&	p	&	OGLE III	&	8389.87538 	&	0.00909 	&	s	&	TESS	\\
	&	2650.75783 	&	0.00482 	&	s	&	OGLE III	&	8390.42113 	&	0.01613 	&	p	&	TESS	\\
	&	2707.96841 	&	0.00334 	&	s	&	OGLE III	&	8392.58274 	&	0.01267 	&	p	&	TESS	\\
	&	2752.76360 	&	0.00303 	&	p	&	OGLE III	&	8393.10787 	&	0.00874 	&	s	&	TESS	\\
	&	2826.70336 	&	0.00298 	&	s	&	OGLE III	&	8492.94966 	&	0.01050 	&	p	&	TESS	\\
	&	2903.88098 	&	0.00245 	&	p	&	OGLE III	&	8493.49392 	&	0.01368 	&	s	&	TESS	\\
	&	3085.76074 	&	0.00537 	&	s	&	OGLE III	&	8496.18907 	&	0.01132 	&	p	&	TESS	\\
	&	3144.59015 	&	0.00272 	&	p	&	OGLE III	&	8497.26820 	&	0.01172 	&	p	&	TESS	\\
	&	3400.40688 	&	0.00288 	&	p	&	OGLE III	&	8499.43080 	&	0.01127 	&	p	&	TESS	\\
	&	3405.26498 	&	0.00697 	&	s	&	OGLE III	&	8502.66542 	&	0.01029 	&	p	&	TESS	\\
	&	3769.55827 	&	0.00324 	&	p	&	OGLE III	&	8505.90916 	&	0.01017 	&	p	&	TESS	\\
	&	3926.60614 	&	0.00568 	&	s	&	OGLE III	&	8509.14666 	&	0.00775 	&	p	&	TESS	\\
	&	4118.19836 	&	0.00343 	&	p	&	OGLE III	&	8510.76999 	&	0.01187 	&	s	&	TESS	\\
	&	4221.27752 	&	0.00468 	&	s	&	OGLE III	&	8511.29412 	&	0.00848 	&	p	&	TESS	\\
	&	4439.30678 	&	0.00530 	&	s	&	OGLE III	&	8511.85083 	&	0.01324 	&	s	&	TESS	\\
	&	4539.14714 	&	0.00384 	&	p	&	OGLE III	&	8512.37044 	&	0.00779 	&	p	&	TESS	\\
	&	4730.19891 	&	0.00198 	&	p	&	OGLE III	&	8514.53386 	&	0.00873 	&	p	&	TESS	\\
	&	4762.04002 	&	0.00496 	&	s	&	OGLE III	&	8661.88088 	&	0.01521 	&	s	&	TESS	\\
	&	4980.61809 	&	0.00234 	&	p	&	OGLE III	&	8663.49891 	&	0.01341 	&	p	&	TESS	\\
	&	5256.93270 	&	0.00367 	&	p	&	OGLE IV	&	8664.57748 	&	0.00953 	&	p	&	TESS	\\
	&	5304.96403 	&	0.00228 	&	s	&	OGLE IV	&	8665.65582 	&	0.01444 	&	p	&	TESS	\\
	&	5500.33045 	&	0.00170 	&	s	&	OGLE IV									\\\hline

\label{tab:min}
\end{longtable}


\end{document}